\documentclass[preprint2,longabstract]{aastex}

\def\red#1 {\textcolor{red}{#1}\ }
\def\green#1 {\textcolor{green}{#1}\ }
\def\blue#1 {\textcolor{blue}{#1}\ }
\def\magenta#1 {\textcolor{magenta}{#1}\ }

\shorttitle{The Distance to NGC4258}
\shortauthors{Argon et al.}

\newcommand\as    {\ifmmode {\rlap.}{''}\! \else ${\rlap.}{''}\!$\fi}
\newcommand\asec  {\ifmmode {\rlap.}^{s}\! \else ${\rlap.}^{s}\!$\fi}
\newcommand\porm  {\ifmmode {~\pm~} \else {$~\pm~$} \fi}
\newcommand\kms   {\ifmmode {{\rm km s}^{-1}}\else{km s$^{-1}$}\fi}
\newcommand\msun  {\ifmmode {{\rm M}_\odot}\else{${\rm M}_\odot$}\fi}
\newcommand\lsun  {\ifmmode {{\rm L}_\odot}\else{${\rm L}_\odot$}\fi}

\newbox\grsign \setbox\grsign=\hbox{$>$} \newdimen\grdimen \grdimen=\ht\grsign
\newbox\laxbox \newbox\gaxbox
\setbox\gaxbox=\hbox{\raise.5ex\hbox{$>$}\llap
     {\lower.5ex\hbox{$\sim$}}}\ht1=\grdimen\dp1=0pt
\setbox\laxbox=\hbox{\raise.5ex\hbox{$<$}\llap
     {\lower.5ex\hbox{$\sim$}}}\ht2=\grdimen\dp2=0pt
\def\gax{\mathrel{\copy\gaxbox}}

\received{2006 December 15}
\begin{document}

\title{Toward a New Geometric Distance to the Active Galaxy NGC\,4258: \\ I. 
VLBI Monitoring of Water Maser Emission}

\author{A. L. Argon, L. J. Greenhill, M. J. Reid, J. M. Moran, \& 
E. M. L. Humphreys}
\affil{Harvard-Smithsonian Center for Astrophysics, 60 Garden Street, 
    Cambridge, MA 02138\\}

\begin{abstract}

{We report a three year, 18 epoch, VLBI monitoring study of 
H$_2$O masers in the sub-parsec, warped, accretion disk within the NGC\,4258
AGN. Our immediate goals are to trace the geometry of the underlying disk, 
track rotation via measurement of proper motion, and ascertain the radii of
masers for which centripetal acceleration may be measured separately. The
monitoring includes $\sim 4\times$ as many epochs,
$\sim 3\times$ denser sampling, and tighter control over sources of systematic error than
earlier VLBI investigations.  Coverage of a $\sim 2400$~km\,s$^{-1}$
bandwidth has also enabled mapping of molecular material $\sim 30\%$ closer to the black hole
than accomplished  previously, which will strengthen geometric and dynamical disk models.  
Through repeated observation we have also measured for the first time a $5\mu$as ($1\sigma$) 
thickness of the maser medium.  Assuming this corresponds to the thickness of the accretion disk, 
hydrostatic equilibrium requires a disk plane temperature of $\approx$ 600 K.  
Our long-term goal is a geometric distance to NGC\,4258 that
is accurate to $\sim 3\%$, a $\sim 2\times$ improvement over the current best
estimate.  A geometric estimate of distance can be compared to
distances obtained from analysis of Cepheid light curves, with the intent to
recalibrate the extragalactic distance scale with reduced systematic uncertainties. This
is the first paper in a series.  We present here VLBI observations, data
reduction, and temporal and spatial characteristics of the maser emission.  Later papers
will report estimation of  orbital acceleration and proper motion, modeling of disk 3-D
geometry and dynamics, and estimation of a `` maser distance.'' Estimation of a ``Cepheid 
distance'' is presented in a parallel paper series.}

\end{abstract}

\keywords{galaxies: individual (NGC,4258) --- galaxies: kinematics and dynamics --- galaxies: nuclei --- galaxies: active --- masers --- techniques: interferometric}

\section{Introduction}

The extragalactic distance scale (EDS) is underpinned by measurements of ``standard candles'' in the 
Large Magellanic Cloud (LMC) and other nearby galaxies \citep{fre01}.  The long-term importance 
of a high-accuracy EDS calibration lies in potential precision estimation of cosmological 
parameters, such as the equation of state for dark energy \citep[e.g.,][]{hu05}.  One means by 
which the EDS calibration might be improved is by anchoring ``standard candle'' calibration to distances 
estimated by geometric techniques.  Towards this end, NGC\,4258---for which a geometric distance has been 
estimated from modelling of the 3-D geometry and dynamics of the accretion disk surrounding the central 
engine \citep{hmg99}---could either replace or be used in conjunction with the LMC as anchor.

In NGC\,4258, accretion disk material emits H$_2$O maser emission in the $6_{16}-5_{23}$ transition 
at 22.235 GHz, which was first reported close to the systemic velocity ($V_{sys}$) by 
\citet{chl84}.  A key discovery was high-velocity Doppler components at $\sim V_{sys} \pm 1000$ km 
s$^{-1}$ \citep{nim93} in NGC\,4258, where $V_{sys} = 472 \pm 4$ km  s$^{-1}$, adopting the Local 
Standard of Rest (LSR) and radio definition of Doppler shift \citep{cwt92}.  An early Very Long Baseline 
Interferometry (VLBI) study \citep{gjm95} suggested that maser emission might originate from a 
rotating disk, as did a parallel investigation of secular velocity drifts for 
the peaks of spectral features, i.e., centripetal accelerations (Haschick, 
Baan, \& Peng 1994; Greenhill et al. 1995b; Nakai et al. 1995; see also Watson 
\& Wallin 1994).  Low-velocity Doppler components (those near 
$V_{sys}$) drifted by $\approx +10$\,km\,s$^{-1}$\,yr$^{-1}$ and high-velocity 
components drifted by $< \pm1$\,km\,s$^{-1}$\,yr$^{-1}$, thus localizing the former to 
the near side of the disk and the latter near the mid-line (i.e., the 
diameter perpendicular to the line of sight).  The sub-parsec disk model was 
confirmed by Very Long Baseline Array (VLBA)\footnote  {The VLBA is a facility of the 
National Radio Astronomy Observatory, which is operated by Associated Universities 
Inc., under cooperative agreement with the National Science Foundation.} 
observations, which traced, most importantly, the disk warp and the rotation curve, 
the latter being Keplerian to better than 1\% (Miyoshi et al. 1995; see also 
Herrnstein et al. 2005).

For a Keplerian system, accurate first-order estimates of geometric 
distance may be obtained relatively simply from the geometry of the warped disk and 
measurements of acceleration or proper motion for low-velocity emission.  To 
date, NGC\,4258 is the only galaxy for which this has been done. 
\citet{hmg99} obtained a distance to NGC\,4258 of $7.2\pm 0.3~(random)\pm 
0.4~(systematic)$ Mpc, where the systematic term is due mostly to a thus far 
weak constraint on orbital eccentricity ($< 0.1$). Our major  aim is to reduce 
both random and  systematic error.  \citet{hmg99} used four epochs of VLBA data 
spanning 3 years to estimate distance.  In contrast, we have 18 epochs over 
a subsequent 3.4 year interval. This facilitates more detailed modeling of disk 
substructure, formal incorporation of orbital eccentricity and periapsis angle 
as parameters, more definitive tracking of disk rotation, and reduction of 
random and systematic error by a factor of at least two. \citet{hag05} discuss 
early progress.

The current distance scale, as characterized by the Hubble constant
(H$_\circ$), may be uncertain by perhaps as little as 10\% \citep{fre01}.  The
most robust estimate depends on measured mean magnitudes of $\sim 800$ Cepheid
variable stars in 31 galaxies within 30\,Mpc, and comparison with the period
luminosity (PL) relation for Cepheids in the Large Magellanic Cloud (LMC).  The
LMC establishes a zero point for calibration of Cepheid light curves in other
galaxies.  Common  secondary distance indicators that are identifiable in the
Hubble Flow (e.g., Tully Fisher relation, type Ia SNe), are calibrated using
distances to Cepheid host galaxies and are used in turn to estimate H$_\circ$. 
High-precision estimates of H$_\circ$ are also available from analysis of Cosmic
Microwave Background (CMB) fluctuations
\citep[e.g.,][]{spe06}.  However, degeneracies in analysis of CMB data do not
allow independent determination of both H$_\circ$ and the equation of state for
dark energy, $w$.  Note that \citet{spe06} assumed a flat universe with $w=-1$ 
and obtained a value of H$_\circ$ in good agreement with \citet{fre01}.

The most substantial problem with the current Cepheid-based calibration of
the EDS is dependence on the LMC as zero point anchor for PL relations. 
Contributing factors include uncertainty in  distance to the LMC, poorly
understood internal galactic structure, and controversy over the impact of the
difference in stellar metallicity between the LMC and HST Cepheid-sample
galaxies.

The adopted uncertainty in the LMC distance by \citet{fre01} is $\sim
5\%$.  However, LMC distance estimates obtained from different calibrations and
analyses over the last decade cover a $\pm 23\%$ range ($\pm 0.50$ mag when
expressed in terms of distance moduli).  This is well beyond uncertainties for
individual measurements \citep[Figure 8 of][]{bmf02a} and likely indicative of
unresolved systematic errors.   Although uncertainties  $\ll 10\%$ have been
obtained from the distributions of distance estimates for different subsamples
of measurements \citep[e.g.,][]{fre01,bmf02b}, in the face of apparently
unresolved systematics, the averages and moments may be statistically problematic.

The uncertain structure of the LMC (e.g., disk eccentricity, out-of-plane
stellar component),  complicates interpretation of distance indicators
\citep[][and references therein]{vdm01, nik04}. For instance, robust estimates
of distance have been obtained for eclipsing binaries, with individual
uncertainties of 2 - 3\% \citep[e.g.,][and references therein]{frg02}. 
However, for the four systems well studied thus far, the range of implied
distances for the LMC center covers 14\% \citep{frg03}.

The impact of metallicity could be as large at $\pm 10\%$ in distance.  Limited
theoretical and empirical understanding has led to long term controversy over the
magnitude and sign of the effect of metal content on Cepheid brightness
\citep{sasselov97, kochanek97, kennicutt98, gro04, sakai04, romaniello06}. Attempts
to make definitive measurements have been complicated by the magnitude
of observable metallicity gradients in well studied galaxies, the sizes of Cepheid
samples, and the effects of reddening and crowding in low-metallicity fields at
smaller galactocentric radii.  Consensus theoretical models have not emerged
because of computational difficulty, with some doubt as to whether or not metal
content alone needs to be considered \citep{fiorentino02}.

Precision estimation of cosmological parameters is best achieved by multiple
approaches.  \citet{hu05} argues that percent level accuracy in H$_\circ$, combined
with measurements of CMB fluctuations, would enable robust constraint on
$w$ independent of those provided by measurements of, e.g., weak lensing and large
scale structure.  \citet{spe06} also demonstrate the impact on estimates of 
curvature by independent constraint on H$_\circ$.  A new ``maser distance'' 
for NGC4258 is a first step, with the immediate goal of $\sim 3\%$.  
This paper is the first in a series. Here we present
the VLBI data and briefly discuss general trends such as time evolution of
position-velocity traces.  Later work will present time-series maser spectroscopy and
modeling of the combined datasets to obtain an improved geometric distance
\citep[cf.][]{hmg99}.  Parallel analysis of recent Cepheid photometry obtained with
the Hubble Space Telescope, beginning with \citet{macri07}, will provide an improved
``Cepheid distance'' \citep[cf.][]{newman01}.  Ultimately, the two
programs will culminate in comparison of maser and Cepheid distances and
re-assessment of the EDS and implications for cosmology.

\section{Observations and Data Reduction}

\subsection{The Observations}

The design of the observing program was intended to enable (1) unambiguous
tracking of evolution in the source spectrum and disk structure, through close
spacing of epochs; (2) monitoring to detect possible transient emission outside the
historically recognized velocity range of the maser, i.e., three 200\,km\,s$^{-1}$-wide
line complexes distributed over $\sim 2000$\,km\,s$^{-1}$~\cite[e.g.,][]{bgm00}; (3) tight
control over sources of systematic error; and (4) relatively uniform sensitivity from 
epoch to epoch.

We observed NGC\,4258 18 times between 1997 March 6 and 2000 August 12,
using the VLBA, the Very  Large Array (VLA), and the Effelsberg 100-m telescope
(EFLS) of the Max-Planck-Institut f\"ur  Radioastronomie (Table 1).  Twelve
``medium-sensitivity'' epochs involved the VLBA alone; the remaining six 
``high-sensitivity'' VLBI epochs were interleaved and involved the VLBA, 
augmented by the phased VLA and EFLS.  

In medium-sensitivity epochs, we recorded eight 16 MHz ($\approx$
216\,km\,s$^{-1}$) basebands with 1-bit sampling. We overlapped seven bands by 15\%,
resulting in observing bandpasses of $\sim 362$ to 1676\,km\,s$^{-1}$ (covering 
low-velocity and red-shifted high-velocity emission) and $\sim -706$ to 608\,km\,s$^{-1}$ 
(covering low-velocity and blue-shifted high-velocity emission), 
depending upon the epoch (Table\,2).  
We tuned the eighth band to enable dual polarization observations of the emission peak 
(spanning the velocity range 362-578 km s$^{-1}$ for red-shifted epochs and 392-608 km s$^{-1}$ for 
blue-shifted epochs), thus increasing 
the signal-to-noise ratio available for tracking atmospheric and instrumental response
fluctuations during calibration.  Because the $\sim 2000$\,km\,s$^{-1}$ bandwidth of the
source was difficult to accommodate instantaneously with VLBI data acquisition systems,
we alternated month-to-month between observing two offset $\sim 1500$\,km\,s$^{-1}$
bands. One covered low-velocity and red-shifted high-velocity maser emission, and the
other covered low-velocity and blue-shifted high-velocity emission. Combining data from
adjacent months, we effectively observed $\sim 2400$\,km\,s$^{-1}$, centered approximately
on the systemic velocity.

In high-sensitivity epochs, we recorded eight 8 MHz ($\approx$
108\,km\,s$^{-1}$) basebands with 2-bit sampling, alternating baseband tuning
over the course of each track to cover the three emission complexes. The
narrower instantaneous bandwidth accommodated the limited IF bandwidth of the VLA
($2\times 50$~MHz).  In the first five high-sensitivity epochs, we recorded both upper 
and lower sidebands to limit the number of baseband mixers required and thereby accommodate a 
reduced complement at Effelsberg.  To avoid band gaps at mixer
frequencies, we overlapped the sidebands of adjacent mixers and obtained
$\approx$ 300\,km\,s$^{-1}$ coverage of low-velocity, red-shifted, and blue-shifted
emission complexes. In the last high-sensitivity epoch, we took advantage of a new
complement of mixers at Effelsberg, recording only upper sidebands with 20\% overlap and
365\,km\,s$^{-1}$ coverage of each emission line complex.

\subsection{Strengths of the Current Study}
 
\noindent
{\it Time Sampling} --- Spacing between high-sensitivity epochs, $\sim
8$ months, was comparable to spacing between epochs in an earlier VLBI study
\citep[][their Table~1]{hmgt05}, but the 1-2 month time interval between 
most medium-sensitivity epochs was comparable to the typical known timescale
for spectral variability \cite[e.g.,][]{bgm00}. A minimum 8 day separation between epochs
also enabled testing for short-term source structure variability. 

\noindent
{\it Bandwidth} --- The high-sensitivity epochs (as in all previous VLBI studies)
focused on the velocity ranges in which \citet{nim93} identified emission. In
contrast, medium sensitivity observations monitored all velocities from  $v_{\rm\tt
LSR}\sim -706$ to 1676\,km\,s$^{-1}$, achieving a detection threshold $\approx 2\times$
below that of earlier broadband ($-$1200 to 2650\,km\,s$^{-1}$) single-dish studies by
\citet{nim93} and \citet{nim95}, scaled to the same channel spacing.  Emission at
``new'' velocities was sought because it would presumably arise in previously
unmapped portions of the accretion disk.  For each epoch, we recorded eight basebands (which were 
correlated in two passes).  Each baseband was divided into 512 spectral channels. 

\noindent
{\it Quality and Consistency} --- The same observing setup and data reduction path
were used for each medium-sensitivity epoch.  A separate common setup and
reduction path was used for high-sensitivity epochs.  Deliberate repetition
(e.g., {\it u,v}-coverage, choice and timing of calibrators, baseband tuning) limited
systematic errors in differential measurements of dynamical quantities (i.e., proper
motion, acceleration).

\noindent
{\it Sensitivity} --- Noise levels ($1\sigma$) in synthesis images were 
2.3 - 4.7 mJy with a 3.0 mJy median for high-sensitivity epochs and 3.6 - 5.8 
mJy with a 4.3 mJy median for medium-sensitivity epochs (Table~1).  Earlier 
studies of low- and high-velocity emission together achieved sensitivities of 
3.0 - 6.5 mJy with a 5.6 mJy median \citep{hmgt05}.

\subsection{Calibration and Imaging}
Calibration and synthesis imaging were performed using
the Astronomical Image Processing System (AIPS) software package 
with largely standard techniques for spectral-line VLBI\footnote 
{AIPS cookbook, chapter~9:  http://www.aoc.nrao.edu/aips/cook.html}.  We discuss
notable elements of the processing below.

\subsubsection{Array and Sky Geometry}
We sought to ensure consistency in the antenna coordinates 
used for each correlation over the 3 years of monitoring.  This was achieved by implementing 
(post-processing) the station positions and (tectonic) velocities estimated by
the United States Naval Observatory (USNO) in solution N9810 (Table~3).  Because the VLA
does not have dual-frequency geodetic VLBI receivers and is absent from solution N9810, we
adopted a position estimated by the NRAO using single-frequency data for epoch 2000.9
and the average velocity for the nearby Pietown and Los Alamos stations (C. Walker,
private communication).  After correction, station positions were accurate to $\approx$ 1
mm for the VLBA and EFLS and $\approx$ 3\,cm for the VLA.

To ensure consistency in the maser position and error budget from epoch to epoch, 
we adopted a maser position obtained from analysis of early VLBI data: 
$\alpha({\rm J2000}) = 12^h 18^m 57\,\asec\, 5046\pm 0\,\asec\, 0003;\ 
\delta({\rm J2000}) = 47^{\circ} 18' 14\,\as\, 303\pm 0\,\as\, 003$ \citep{hmgt05}. 
This was used in correlation of all but the first three epochs, for which we applied a 
frequency-dependent shift to the measured maser spot positions, after \citet{hmgt05}.  
The shift corresponded to an error of $-0\,\asec\, 0054$ and $0\,\as\, 033$ in
right ascension and declination, respectively, in the {\it a priori} position.  
{\it A priori} calibrator positions were 
accurate to $\la 1$\,mas, and no corrections were made.  {\it A priori} earth orientation
parameters  were compared to ``final'' values published by the USNO and found to be
accuarate to $< 1$\,mas (pole position in X and Y) and $<0.1$\,ms (UT1-UTC) for each
epoch.  We discuss the impact of these uncertainties and limits in $\S 2.5$ (see also
Table~4).

\subsubsection{Amplitude Calibration}
To calibrate VLBA fringe amplitudes, we applied observatory-standard antenna gain
curves and time-series system temperature measurements.  An $\approx40$\% 
depression in the system temperatures of the red-shifted high-velocity basebands 
in high-sensitivity epochs rendered some of these calibrations suspect.  
Such large changes in system temperature is not likely and is probably due to a 
measurement problem.  To compensate, we computed a system temperature time-series 
characteristic of the several low-velocity basebands and applied it to all basebands.

Accurate system temperature time-series were typically not available for the 
VLA and EFLS.  For these two stations, we calibrated amplitudes through a
comparison of fringe amplitudes on 4C39.25 among VLBA stations (after
calibration) and fringe amplitudes for baselines including the VLA or
EFLS.  Time sampling of the resulting calibration was relatively coarse,
but we anticipated later self-calibration would enable refinement.

We note that in the three cases for which we made nonstandard amplitude corrections
(i.e., VLA, EFLS, and the red-shifted high-velocity VLBA basebands in high-sensitivity
epochs) amplitude errors for some basebands could survive self-calibration, because
gain corrections are calculated for emission in just one (low-velocity) baseband. 
Though a matter of concern, amplitude errors on the order of a few tens of percent on 
a small fraction of baselines are anticipated to have a relatively small impact on 
astrometry for point-like sources, i.e., maser spots.

We expect the overall flux scale for a given epoch to be accurate to $\approx 30\%$ 
for medium-sensitivity epochs and $\approx 50\%$ for high-sensitivity epochs.  
However, when comparing the overall flux scale {\it between or among} epochs, 
we expect an accuracy of $\approx 15\%$ for medium-sensitivity epochs and 
$\approx 25\%$ for high-sensitivity epochs. 

\subsubsection{Global Fringe fitting}
The most demanding element of the global fringe fitting process was the accurate measurement of
the  slowly time variable electronic phase offsets among basebands, using observations of
quasars made approximately every hour for that purpose.  Specifically, the wide spacing in
frequency among bands increased susceptibility to (nonphysical) phase wraps entering the
interpolation of phase offsets between adjacent calibrator scans.  Because the phase wraps were
readily detected, straight-forward editing of single-band delay and rate data enabled
removal and estimation of offsets with accuracies on the order of $10^\circ$ or $\sim
3\%$ of a fringe on the sky.

\subsubsection{Self-calibration}
We applied self-calibration solutions to the data in each baseband to correct for fluctuations 
in fringe amplitude and phase.  These fluctuations arose from instabilities in the atmosphere 
and in the frequency standards at the stations and from antenna gain changes.  
For each epoch, we identified a relatively unblended Doppler component ($\gax$ 1\,Jy) 
to provide the reference signal.  Images of the reference component achieved dynamic 
ranges of 40-300, defined with respect to the $5\sigma$ noise level or strongest artifact.  

We observed unusually large artifacts in images of high-velocity red-shifted emission
for {\it a few} high-sensitivity epochs.  Dynamic ranges were reduced by a factor of
$\sim 4$.  We were unable to identify a cause and adopted an empirical strategy to
correct the problem.  For these epochs, we performed a second self-calibration on the
strongest observed red-shifted Doppler component.  Though typically only a few tenths of a
Jy, this secondary reference was detectable because the initial self-calibration extended
the phase coherence of the data.   We applied the additional, slowly varying amplitude and
phase corrections to the data in all red-shifted basebands.  To recover relative
position information (with respect to the primary reference emission), we applied the
angular offset between reference features, measured prior to the second
self-calibration.  

Secondary self-calibration of the blue-shifted high-velocity emission was not 
possible because the peak flux density was too low ($< 100$\,mJy).  We tested two secondary 
calibrations for the ``blue data''.  The first was achieved by applying the ``red data'' 
self-calibration corrections to the ``blue data'' and the second by applying the complex 
conjugate of the ``red data'' corrections to the ``blue data''.  In the first case we implicitly 
modeled the unexplained errors as phase-only, while in the second case as delay-only.  Neither 
approach improved the signal-to-noise in images of blue-shifted emission, and was therefore 
not adopted.  Some losses in imaging blue-shifted emission remain possible.  

\subsubsection{Imaging and Deconvolution}
We imaged and CLEANed a $25.6\times 25.6$\,mas field (in medium-sensitivity epochs) or
a $30.72\times 30.72$\,mas field (in high-sensitivity epochs) over the inner 96\% of each
bandpass. The known angular extent of the maser source is $\sim1\times16$\,mas. 
Synthesized beam sizes and typical noise levels for individual spectral channels 
are given in Table~1.  We used a hybrid weighting scheme midway between uniform
and natural weighting (ROBUST=0 in the AIPS task IMAGR), in order to balance the
desire for a small beam measured at the half-power points and low
rms noise.  In five of the six high-sensitivity epochs, we extended imaging
outside the inner 96\% of the bandpasses, because a blue-shifted Doppler component 
had been found in the medium-sensitivity epochs.  The anticipated emission 
was detected in one epoch (BM081A). 

\subsection{Identification of Maser Spots and Continuum Emission}
We fit a 2-D Gaussian brightness distribution to all peaks in 
channel maps and required that they satisfy the following requirements, in order to 
be judged {\it bona fide} maser ``spots.''

\begin{itemize}
\item{Emission centroids lie within one beamwidth over at least
three adjacent channels at or above the 6$\sigma$ level.}

\item{Fitted peak intensities exceed the intensity of imaging artifacts, i.e., 
intensities must exceed the local peak intensity divided by the dynamic range. We note 
that this dynamic range cut was only relevant to channels in high-sensitivity 
epochs with strong emission features and candidate secondary features.}

\item{Half-power major and minor axes are both smaller than twice the
major and minor axes of the convolving beam.  In cases where breadth appeared
consistent with blending of individual maser spots, we attempted to fit
multi-component Gaussian models.}

\end{itemize}

In each epoch, there are on the order of $10^7$ independent pixels among the image
cubes.  The probability that one or more will exceed  6$\sigma$, in the absence of a
signal, is $\approx 10\%$ \citep[e.g.,][Fig. 9.5]{tms01}.  By requiring 
three adjacent channels to satisfy this criterion, with a constraint on
position, the chances of false detection are on the order of $10^{-7}$ (where we
have accounted for correlation among channels consistent with the FX architecture and
time-windowing of data in the VLBA correlator).  

Where we had {\it a priori} information as to the location in velocity and position 
of a maser, we relaxed the 6$\sigma$ selection criterion.  There were three instances. 

\begin{itemize}
\item{High-velocity features ($>3.5\sigma$) were accepted if 
emission at corresponding positions and velocities had been detected above the 
formal detection threshold in other epochs.  The probability 
of false detection of a $3.5\sigma$ emission peak over one independent point (one beam) 
is $<10^{-3}$.  Blue-shifted Doppler components falling into this 
category were found at $-285$ km s$^{-1}$ (BM081A), $-367$ km s$^{-1}$ (BM112G), 
$-373$ km s$^{-1}$ (BM081B and BM112C), and $-515$ km s$^{-1}$ 
(BM112C, BM112E, and BM112H).  Red-shifted Doppler components falling into this category 
were found at 1566 km s$^{-1}$ (BM112F and BM112M).}

\item{High-velocity blue-shifted emission ($>2\sigma$) was accepted in spectral 
channels adjacent to {\it already identified} blue-shifted features, if 
in the same epoch and at the same angular position to within a beam width.  
Blue-shifted emission was identified in this way at $-285$ km s$^{-1}$ (BM112C, BM112E, BM112G, 
BM112L, BM112N, and BM112P), $-367$ km s$^{-1}$ (BM112C), $-373$ km s$^{-1}$ (BM056C, 
BM112E, BM112G, BM112H, and BM112L), $-435$ km s$^{-1}$ (BM056C and BM081A), $-441$ km s$^{-1}$ 
(BM056C), and $-515$ km s$^{-1}$ (BM112G).}  The motivation for this extension of our detection 
limit was better establishment of line shape for weak features.  It was not used to increase 
the number of features.

\item{High-velocity features ($>3.5\sigma$) outside the nominal range 
\citep[i.e.,][]{nim93} were accepted, if single-dish study separately identified the
Doppler components.  In this way, we detected the 1647\,km\,s$^{-1}$
component first reported by \citet{mmk05} for epoch 2003.75.  Extrapolating from the
positions of other  red-shifted Doppler components and adopting a Keplerian rotation curve
for the ``maser disk,'' we confined the anticipated position of the 
1647\,km\,s$^{-1}$ emission to within 1 beam $\times$ 3 channels.   Emission was detected 
in BM112M at 5$\sigma$ in the line-center channel and 4$\sigma$ in adjacent
channels. Marginal detections were also found in two adjacent epochs, 
i.e., emission at 2.4 and 3.0$\sigma$ in two contiguous spectral channels for BM112K 
and four $2.8\sigma-3.3\sigma$ points in adjacent spectral channels for BM112O. }
\end{itemize}

In addition to identifying maser emission, we searched for continuum emission 
at all epochs within 12.5\,mas of the low-velocity reference emission.  
Continuum {\it u,v}-data were constructed by frequency averaging each baseband prior to imaging.  
In the case of BG107 and all medium-sensitivity epochs, we discarded 
the outer 2\% of each of the red- or blue-shifted basebands and obtained an effective 
average of 1960 channels (30.6 and 61.3 MHz for BG107 and medium-sensitivity epochs, 
respectively).  In the other high-sensitivity epochs, we discarded 
the outer 2\% of each of three red- or blue-shifted basebands and averaged 1470 channels 
(23.0 MHz).  Fewer channels were retained in these epochs, 
because of extensive overlap of sidebands for adjacent baseband mixers.


Finally, we performed a separate search for maser emission that might be
associated with jet activity in NGC\,4258, well downstream from the central engine. In 
the most sensitive of our high-sensitivity observations (BM112H), we boxcar
averaged the {\it u,v}-data over four channels to improve sensitivity and imaged a
$30.72 \times 491.52$ mas ($512 \times 8192$ pixels, E-W$\times$N-S) strip 
centered on the reference position.  No emission peaks
exceeded a 6$\sigma$ detection threshold (13\,mJy).  

\subsection{The Error Budget}
For each Gaussian fit, the random measurement error in position was estimated as 
being the larger of the fitting error or the beam full-width at half-maximum (FWHM) 
divided by twice the peak signal-to-noise ratio ($S/N$).  We estimated the systematic 
position error from two components:  (1) errors that scale with difference in sky frequency 
between a given channel and the reference (i.e., $\nu - \nu_{ref}$), and (2) 
errors that scale with the difference in baseband frequencies of a given channel and
the reference (i.e., [$\nu - \nu_\circ] - [\nu_{ref} - \nu_{ref_\circ}]$, where $\nu_\circ$ 
and $\nu_{ref_\circ}$ are the band-center frequencies).  The maximum difference in
sky frequencies is  $\approx$  75 MHz for NGC\,4258; the maximum difference in
baseband frequencies is $<8$ MHz.   Errors in the absolute
position of the reference and calibrators, or errors in zenith
propagation delay and baseline coordinates give rise to
the first type of systematic error.   Uncertainty in station clocks give rise to
the second type. 

A consistency check of the high-accuracy ($\sim 3$ mas) maser 
reference position of \citet{tms01} can be obtained from our continuum imaging of BM081B.  
In this epoch, we imaged with a less accurate ($\sim 60$ mas) {\it a priori} maser 
reference position, which led to a substantial difference between red- and blue-shifted
position offsets, due to the large frequency difference between red- and 
blue-shifted continuum ``bands'' (see Table 4).  
If we assume that this misalignment of red- and blue-shifted continuum 
is entirely due to an inaccurate {\it a priori} maser reference position, we obtain a new 
reference position that is offset from \citet{tms01} by $0\asec 001$ in right ascension 
and $0\as 01$ in declination, well within our $1.5\sigma$ combined errors (ours plus 
Herrnstein's, added in quadrature).  We note that all channel and continuum peak positions in 
this epoch were corrected, post-imaging, for the lower-accuracy {\it a priori} maser 
reference position.

Table~4 gives the magnitudes of random and systematic position  error.  Random
error dominates the error budget for weak masers,  whether they be at low- or
high-velocities. For instance, blue-shifted masers are dominated by random error. 
Systematic error dominates red-shifted masers stronger than $\approx$ 0.1\,Jy and
low-velocity masers stronger than $\approx$ 0.5\,Jy.  The largest contribution to systematic
error arises from  uncertainty in the absolute position of the reference emission, 
causing $\approx$ 10 $\mu$as error across 75 MHz, the typical frequency separation
between low and high-velocity masers.  For a given Doppler component, the
error repeats from epoch to epoch, with second-order impact (because reference
emission velocities differ by $<4$\,MHz among epochs). However, a 10 $\mu$as error 
is comparable to upper limits on disk thickness and the hydrostatic scale height
\citep{mgh95}, with implications discussed in a later paper. 

\subsection{Relativistic Velocity Assignments}
Velocities in the LSR reference frame, assuming the radio definition of
Doppler shift, were assigned to each channel.  The choices of reference frame and
definition were made for consistency with previous work \citep[e.g.,][]{hmgt05}.
However, the first-order (radio definition) and full relativistic treatment of velocity differ by as
much as 4\,km\,s$^{-1}$ for the most red-shifted emission.  Because this is 
comparable in magnitude to important physical effects (e.g., distortion of the rotation
curve by disk warping), we have adopted relativistic velocities for purposes of disk
modeling.  Both first-order and relativistic velocities are presented with the
fitted maser spot parameters in Table~5.

In order to convert classical to relativistic velocities, we 
first reconstructed the actual sky frequency of a feature observed in the earth's 
topocentric frame at 12$^h$ UT for the date of observation.  This involved 
removing the LSR correction to obtain the classical topocentric velocity and converting
to the observed frequency according to the classical radio Doppler formula
$$x = {\nu_{\rm obs}\over \nu_{\rm 0}} = 1 - {v_{\rm ctop}\over c},$$
where $\nu_{\rm obs}$ is the observed sky frequency, $\nu_{\rm 0}$ the 
frequency in the emitter's rest frame, $v_{\rm ctop}$ is the classical velocity 
in the topocentric frame, and $c$ is the speed of light.  
We then converted to a relativistic topocentric velocity, 
$${v_{\rm rtop}\over c} = {(1-x^2)\over (1+x^2)} .$$
Finally, we reintroduced the LSR correction to obtain a relativistic LSR 
velocity,
$$v_{\rm r{\tt LSR}} = {v_{\rm rtop} + \Delta v_{\rm\tt LSR}\over 1 + 
(v_{\rm rtop}\ \Delta v_{\rm\tt LSR})/c^2}.$$

\noindent
After velocity correction, all positions were re-referenced to $v_{\rm  rLSR}$ =
510.0\,km\,s$^{-1}$, corresponding to the classical LSR velocity  $v_{\rm cLSR} = 509.6$
km s$^{-1}$.  We note that in all previous NGC\,4258 studies \citep[e.g.,][]{hmg99}, 
the quoted velocities have been classically defined and conversions performed during
accretion disk modeling made the analyses relativistically correct.  Those 
conversions included a general relativistic correction, which for a black hole of mass
$3.8\times 10^7 \msun$ was $\approx 4$\,km\,s$^{-1}$  at the disk inner radius and
2 km s$^{-1}$ at the outer radius.  We have not applied a general relativistic
correction to the data presented here because the correction is model dependent, and 
its discussion is reserved for a later paper.

\section{Results}
\subsection{Data Products}

\noindent
{\it ~Maser Spectra} --- Spectra of low- and high-velocity masers are 
shown in Figures~1-4.  These spectra were constructed from fitted maser spot
flux densities.  An average spectrum for all epochs is shown in Figure~5.  Channels
without detectable emission are not shown.\\

\noindent
{\it ~Maser Maps} --- We present the full set of Gaussian fits
for detected maser spots in Table 5.  
Note that spots that lie in the overlap between adjacent basebands 
appear twice.  Quoted position uncertainties are fitted random error.  
The true random error is estimated to be the maximum of the fitted 
random error and the random error computed according to Table 4.  
The sky distribution and related statistics of the masers are shown in 
Figures 6-9; the position along the major axis of the disk 
versus LSR velocity are shown in Figures 10-13.  We define the 
impact parameter to be $b = \pm (\Delta x^2 + \Delta y^2)^{1/2},$ 
where $\Delta x$ and  $\Delta y$ are the east-west and north-south offsets from 
the emission at 510.0\,km\,s$^{-1}$ (relativistic), respectively, and the sign of
{\it b} is taken from the sign of $\Delta x$ (which is feasible because the
disk is elongated chiefly east-west).  The formal uncertainty in impact parameter
($\sigma_b$) is given by
$$\sigma_b = \Bigl[\ { {(\Delta x\ \sigma_x)^2 + 
(\Delta y\ \sigma_y)^2}\over {\Delta x^2 + \Delta y^2}}\ \Bigr]^{1/2},$$
where $\sigma_x$ and $\sigma_y$ are the random components of uncertainty 
in east and  north offsets, respectively.   
\\ 

\noindent
{\it Continuum Maps} --- We detected continuum emission $>3\sigma$  
in fifteen epochs and averaged deconvolved images (with uniform 
weighting) to obtain a peak intensity of 1.1 mJy ($23\sigma$) (Figure~14).  
Nondetection for the remaining three epochs may be attributed to known time 
variability of the source \citep{her98}.  In Table~6, we list (angle 
integrated) flux densities and centroid positions.  Figure~15 shows a 
plot of flux density versus time.  Our measurements are 
consistent with the hypothesis that the emission traces a jet extending outward 
from the central engine, along the spin axis of the accretion disk \citep{her98}.

\subsection{Position Shifts}
Visual comparison shows that the position-velocity loci for low- and red-shifted
high-velocity emission are offset at a few epochs by up to $\sim 30\mu$as from a weighted
average locus computed from the medium-sensitivity epochs.  The shifts do not appear to
be frequency dependent.  We attribute them to residual phase offsets 
between basebands correlated in separate passes (as were low and high-velocity bands),
uncertainty in the post self-calibration registration that was required for
particular high-sensitivity epochs, and unmodeled changes in the structure of the
reference emission from epoch to epoch.  

We computed corrections for this effect as follows.  First, we constructed variance-weighted 
average {\it x} and {\it y} positions for each channel from all medium-sensitivity 
epochs.  (Weighting depended upon random measurement error alone.)   Next, we downweighted 
outliers according to the empirical scheme below and computed new variance-weighted 
averages.  To downweight outlying samples, we multiplied measurement errors by an 
empirical factor that increased with offset: $\exp({\Delta\over 4\sigma})^2$, 
where $\Delta$ is the offset in $x$ or $y$ from the weighted average and 
$\sigma$ is the unmodified measurement uncertainty.  For instance, the measurement 
error for points that deviated by $6\sigma$ is increased by a factor of 10, whereas 
the measurement error for points that deviated by $1\sigma$ is increased by a factor of 1.06.  
Finally, we obtained offset corrections for individual epochs (both high- and 
medium-sensitivity) by minimizing the sum of the squares of the offsets from the 
variance-weighted averages above.  Figure 16 shows the observed and corrected
offsets for red-shifted emission in observation BM112H, an epoch with one of the 
largest shifts.  Table~7 quantifies the computed shifts for each epoch, which we 
applied to the data.  However, we note that Table 5 contains uncorrected 
maser spot positions. In addition, no corrections have been computed for 
blue-shifted emission because it is too sparse to enable comparisons.

\section{The Sky Distribution of Maser Emission}
Maser emission in NGC\,4258 traces a thin, slightly warped, 
nearly edge-on disk \citep{mmh95,her98,hmgt05}. The new VLBI data 
(Figure~6) trace this structure more fully, due to both emission time variability 
(i.e., development of new Doppler components) and broader observing bandwidth (e.g., 
velocities above 1500\,km\,s$^{-1}$).  The flaring (in north offset range) of low-velocity 
maser spots at the east offset edges of the plot is an artifact and is due to the fact 
that the weaker spots, which have lower positional accuracy, are preferentially 
located away from the middle of the low-velocity spectrum (Figure~1).  
Crowding of plot symbols also obscures the relative breadths of the distributions 
of maser spots with different flux densities.  To assess the statistics of the 
distribution of maser spots, we have computed perpendicular offsets of low-velocity 
masers from a boxcar-smoothed (30 km s$^{-1}$ width), weighted average locus of 
emission observed in the medium-sensitivity epochs.  For spots $<250$ mJy, the 
distribution is Gaussian and is consistent with 
measurement noise (Figure 8, top panel).  For spots $>250$ mJy, the distribution 
exhibits significant non-Gaussian wings and tail (Figure 8, bottom panel), which is a 
result of source structure on scales of order 10 $\mu$as.  If we restrict the 
velocity range under consideration, the distribution of strong spots may be 
related to accretion disk thickness.  The inclination and position angle warp 
of the disk combine to create a concavity on the near side, the bottom of which 
is tangent to our line of sight (i.e., the disk is seen edge-on at this point).  
Maser spots in the velocity range 485-510 km s$^{-1}$ lie in this part of the 
disk \citep[their Figure 14]{hmgt05}.  Previously measured accelerations for 
masers in this velocity range are 8-10 km s$^{-1}$ yr$^{-1}$, which suggests a 
radius of 3.7-4.1 mas and from the warp model we infer a projected range vertical 
offset of only 3 $\mu$as.  The observed distribution of maser spot position 
offsets from the mean locus is nearly Gaussian with $\sigma$ = 5.1 $\mu$as 
(12 $\mu$as at full-width half-maximum).  Because {\it most of} the maser spot positions in 
this velocity range are measured with $<3 \mu$as accuracy, we propose that 
this Gaussian distribution largely represents a true distribution about the 
local midplane of the disk (i.e., the thickness of warm molecular material.)  
See next section for interpretation.
  
\section{Notes on Source Structure and Evolution}

The long-term importance of VLBI monitoring of the NGC\,4258 H$_2$O maser includes 
contributions to estimation of a geometric distance and its application to
high-accuracy measurement of the Hubble constant.  Estimation of a distance requires 
additional quantification of centripetal accelerations for disk material and modeling of
3-D disk structure and dynamics.  These will be treated in follow-on papers.  However,
we can make several observations based on the time-series VLBI data alone. 

{\it Accretion disk thickness.}--- In prior analyses, scatter in low-velocity maser 
spot positions (north-south) have been used to estimate an upper limit on the 
thickness of the underlying accretion disk.  \citet{mgh95} obtained a limit of 
$\sim 10 \mu$as by focusing on strong maser features toward the middle of the 
spectrum for a VLBI epoch in 1994.  We have combined the results for all 18 VLBI 
epochs to obtain a time-average sample of emission for which the width of the 
distribution ($1\sigma$) is $5.1 \mu$as.  This may be the first directly measured 
thickness for an accretion disk around any black hole.  
In hydrostatic equilibrium the accretion disk is expected to have a Gaussian 
distribution of density as a function of height with $\sigma = r\ c_s/v_{\phi}$, 
where $r$ is the radius, $v_{\phi}$ is the orbital velocity, and $c_s$ is the sound 
speed \citep{fkr02}.  Note that $v_{\phi}/r = (GM/r^3)^{1/2} = \omega$, which is the 
slope of the impact parameter versus velocity curve of Figure 10 and equals 
$2.67\times 10^{-10}$ s$^{-1}$.  Hence, for $\sigma = 5.1\ \mu$as or 
$5.5\times 10^{14}$ cm, we obtain $c_s = 1.5$ km s$^{-1}$, corresponding to a 
temperature of about 600 K (about half the limit given by \citet{mgh95}).  
This temperature is consistent with 
the 400-1000\,K range that is believed necessary for maser action from H$_2$O 
\citep{e92}.  Furthermore, the inferred sound speed is comparable to 
the line width of individual maser Doppler components seen in spectra, which is 
consistent with maser saturation.  We note that in our discussion, we have assumed that 
the maser distribution samples the full thickness of the accretion disk, rather than 
a surface layer, whose depth could be well below that of the physical disk.  This 
assumption is consistent with the disk models of \citet{nm95} for radiatively efficient 
accretion.  A more detailed inspection of residuals from models 
fit to maser spot distributions (e.g., residuals arising from an antisymmetric offset perpendicular 
to the disk plane, caused by irradiation of ``upper'' and ``lower'' surfaces of 
the disk for high-velocity red- and blue-shifted masers) is necessary for further 
comment.  We also note that the measured thickness places a limit on the magnetic 
pressure.  The magnetic field strength must be less than about 100 mG, which 
is about the current limit from the non detection of the Zeeman effect \citep{mmk05}.

{\it Highly red-shifted emission.}--- Red-shifted emission at velocities  
$\gg 1460$\,km\,s$^{-1}$ has been mapped here for the first time.  Emission at
1647\,km\,s$^{-1}$ was first reported by \citet{mmk05} in a single-dish spectrum, while
the 1566\,km\,s$^{-1}$ Doppler component was previously unknown.  The detection of these 
very high velocity emissions enables the mid-line of the disk, along which the rotation
curve is apparent, to be traced to about 30\% smaller radii ($\approx 0.11$~pc) than 
before.  This innermost radius is now also $\approx 20\%$ smaller
than that for the low-velocity emission.  Perhaps most importantly, the radius of the
1647\,km\,s$^{-1}$ emission is small enough that curvature in the disk is directly
observable, which will enble more robust geometric modeling of the warp 
\citep[cf.][]{mmh95, her98}.

The low-velocity emission was shown by \citet{hmgt05} to originate near
the bottom of a concave depression on the near side of the disk, where our line of sight
is tangent to the disk.  We note that in contrast, the 1566\,km\,s$^{-1}$ component lies
near the topmost portion of the spine of the disk projected on the sky.
For an ``alien'' observer viewing the disk edge-on along a line of sight close 
to our sky plane, the material 
we perceive at 1566\,km\,s$^{-1}$ would probably be responsible for strong low-velocity 
emission.  Conceivably, the intensity of that emission would be boosted by seed photons from the
northern lobe of the radio jet, much as the low-velocity emission visible to earthbound
observers is believed to be boosted by the southern lobe of the jet \citep{hmg97}.  The resulting
(competitive) beaming of maser energy along a radial path may be responsible for the
low intensity of the 1566\,km\,s$^{-1}$ emission, for which the dominant gain path is orthogonal.

{\it Source Symmetry.}--- Early VLBI maps of the maser emission
exhibited asymmetry, wherein the blue-shifted high-velocity emission spanned a
truncated range of radii compared to red-shifted emission \citep{mmh95}.  It was unclear 
whether this was a reflection of limited sensitivity (since the
blue-shifted emission is quite weak) or a ramification of unmodeled disk structure. 
Detection of emission over a broader range of radius (i.e., velocity) has been
accomplished since then. \citet{nim95} first observed a broader
range via single-dish spectroscopy, detecting emission near $-300$ 
and $-500$\,km\,s$^{-1}$ one time each, in $\approx 20$ epochs spaced over $\approx
200$~days.   In their Figure~3, \citet{hmgt05} also show data derived from a VLBI detection
of blue-shifted emission near $-300$ and $-500$\,km\,s$^{-1}$ at one epoch in 1996 September.
With new detections shown in our Figure~13 at $\approx -285$\,km\,s$^{-1}$ (7 epochs over 2-3
years) and $\approx -515$\,km\,s$^{-1}$ (2 epochs over 8 days), we confirm the
distribution of blue-shifted high-velocity emission over a 
$\approx 0.1$~pc range of radius. In light of the very high velocity red-shifted emission
reported here (which shows that warm molecular material extends to relatively small
radii), and assuming mirror symmetry of the maser disk, we hypothesize there may be
still higher velocity blue-shifted emission that could be detected in the vicinity of 
$-600$\,km\,s$^{-1}$. However, any such emission is likely to be weak, perhaps on the order 
of 1\,mJy, based on line ratios among known Doppler components.

{\it Envelope of low-velocity emission.}---Our long term monitoring of
velocities between the high and low-velocity emission complexes has not resulted in detected 
intermediate-velocity emission.  Nondetection here and in episodic
single-dish spectroscopy of comparable sensitivity \citep[e.g.,][]{mmk05} may
exclude the possibility of long gain paths at disk azimuth angles more than
$\approx 0.1$\,rad from the emission close to the systemic velocity.   This contrasts with
what has been found in the NGC\,1068 H$_2$O maser \citep{gg97}, wherein the
locus of ``low-velocity'' maser emission extends in a ring over a full quadrant on the
receding side of the disk.  However, in the angular extent of high
brightness temperature continuum emission that can provide seed photons for
amplification differs.  In NGC\,4258, they arise from a compact jet core \citep{hmg97,hmgt05},
while in NGC\,1068, they are believed to arise from an extended region that more or less
fills the volume inside the innermost maser orbit \citep{gallimore01}--though the absence of 
seeded emission in the adjacent (approaching) quadrant is difficult to explain.

Though centripetal acceleration is responsible for the observed secular drift of
individual Doppler component velocities, the overall range of emission does 
not change discernably in angle or velocity.   In our time
monitoring, we detected low-velocity emission from $\approx 380$ to 580\,km\,s$^{-1}$. 
This range is comparable to the ranges observed in high-sensitivity spectroscopic studies over
the period  1992-2005 \citep[e.g.,][]{ghb95,nim95,ysh05,mmk05}. Any secular change of the
upper and lower limits of the envelope appears to be much smaller than the
10\,km\,s$^{-1}$\,yr$^{-1}$ drift of individual Doppler components.

{\it No maser spots seen away from the disk.}---In contrast to maser emission 
from the Circinus galaxy \citep{gre03} and NGC\,1068 \citep{gbo96, gallimore01}, 
we found no Doppler components more than a beamwidth ($\approx$ 1 mas) from the 
locus of disk emission. Our detection limit was 13 mJy ($6\sigma$) for
the highest sensitivity observation, BM112H.  In Circinus, the maser population not
associated with the disk is believed to trace a bipolar, wide-angle outflow that 
contains clumps moving away from the central engine, at radii as small as 
$\approx 0.1$\,pc.  In NGC\,1068, the non-disk emission lies $\gg1$\,pc 
downstream in the narrow-line region, along the jet axis. The absence of emission
associated with wind or jet emission in NGC\,4258 (see $\S 2.4$) 
may be a consequence of a lower 
luminosity central engine and  more tightly bound disk material (i.e., smaller
disk mass compared to the central engine mass).

\par
{\it Absorption signatures.}---\citet{ww94} hypothesized that thermalized
water molecules in disks could absorb maser emission arising at smaller radii; 
the observational signature would be troughs in spectra near systemic 
velocities.  A number of spectra for NGC\,4258 that are in the literature do 
exhibit dips in the vicinity of the systemic velocity, including time averages of 
many spectra collected over periods of months or years \citep[see][and 
references therein, as well as our Figures~1,5]{bgm00}.  The cumulative record of 
VLBI observations does not demonstrate notable structure in the vicinity of the 
systemic velocity, which is consistent with expectations.  
However, evidence thus far has been anecdotal (e.g., compare 
spectra for epochs 2000.08 and 2000.61).  
Systematic analysis of spectra obtained throughout 1985 and 1986 demonstrated that 
troughs in spectra exhibited secular drift similar to that of neighboring emission
components \citep{ghb95}.  A physical model of the NGC\,4258 warped accretion 
disk presents additional hurdles to the hypothesis.  In the case of NGC\,4258, the 
vertical deviation of the disk from a plane displaces material at large radii
from the line of sight \citep{hmgt05}, and the velocity at which the longest
coherence lengths occur (overlapping the rest frame transitions of inverted and thermalized 
material) is displaced $\sim 20$\,km\,s$^{-1}$ redward from the systemic velocity.  
Moreover, the model of \citet{nm95} predicts that material at sufficiently 
large radii is atomic, and thus cannot form an absorbing layer.


\section{Conclusion}
We have reported a four year time-series VLBI study of the H$_2$O maser in
NGC\,4258, with quadruple the number of epochs than earlier work.  Overall source 
structure during the four years of our study resembled what has previously been 
fitted to an inclined, warped, Keplerian disk model.  
Frequent sampling over years enabled compilation of a more complete census 
of time variable Doppler components and broad observing bandwidth enabled 
detection and mapping of material with larger orbital velocities than before.  
The persistence and symmetry of source structure is now more clear, and 
the accuracy of position measurement has been improved.  Detailed analysis 
of the distribution of maser spots perpendicular to the local disk plane, 
interpreted in the context of hydrostatic support, suggests a disk thickness 
($\sigma$ = $5.5\times 10^{14}$ cm) and gas temperature of about 600 K.  
The new catalog of maser spot positions and line-of-sight
velocities will enable construction of a more robust (3-D) dynamical disk
model and more accurate measurement of maser centripetal acceleration and
proper motion.  Together, these will enable estimation of a geometric
distance with smaller random and systematic uncertainty \citep[cf.][]{hmg99}.  
Follow-up papers will treat the measurement of acceleration,
modeling, and comparison of a final ``maser distance'' to a new four-color
Cepheid distance that enables recalibration of distance indicators and
ultimately H$_\circ$.

{\acknowledgements}
We are grateful to B. Clark, M. Claussen, J. Wrobel, and the NRAO data analysts
for their help in queuing our dynamic schedules on the VLBA.  We thank D. Graham, 
A. Kraus, and C. Henkel for assistance in operating the Effelsberg telescope and 
M. Eubanks, A. Fey, J. Herrnstein, A. Trotter, and C. Walker for helpful discussions and
comments. This work was partially supported by HST Grants GO-09810 and HST-GO-10399, and
by NASA Grant NAG5-10311.

\clearpage

\begin{deluxetable}{clllccccc}
\tabletypesize{\scriptsize}
\rotate
\tablewidth{0pt}
\tablecaption{The Observations}
\tablehead{
\colhead{Experiment Code}                             & 
\colhead{Date}                                        &
\colhead{Antennas\tablenotemark{a}}                   & 
\colhead{Synthesized Beam\tablenotemark{b}}           &
\colhead{Sensitivity\tablenotemark{c}}                &
\multicolumn{3}{c}{Coverage\tablenotemark{d}}         & 
\colhead{Comments}\\
&
&
&
\colhead{(mas$\times$mas, deg)}&
\colhead{(mJy)}&
\multicolumn{3}{c}{$\hrulefill$}&
\\
&
&
&
&
&
\colhead{L}&
\colhead{R}&
\colhead{B}&
}

\startdata
BM056C\dotfill& 1997 March 06 (1997.18)    & VLBA, VLA, EFLS& 0.65$\times$0.53, \phantom{$+$}12.5& 3.0 & x& x& x&              \\  
BM081A\dotfill& 1997 October 01 (1997.75)  & VLBA, VLA, EFLS& 0.59$\times$0.35, \phantom{1}$-$7.2& 2.7 & x& x& x&              \\
BM081B\dotfill& 1998 January 27 (1998.07)  & VLBA, VLA, EFLS& 0.60$\times$0.42, \phantom{1$+$}5.1& 3.2 & x& x& x&              \\
BM112A\dotfill& 1998 September 05 (1998.68)& VLBA, VLA, EFLS& 0.67$\times$0.49, \phantom{1$+$}8.5& 3.0 & x& x& x&              \\
BM112B\dotfill& 1998 October 18 (1998.80)  & VLBA           & 0.52$\times$0.36, $-$18.0          & 4.2 & x& x&  &              \\
BM112C\dotfill& 1998 November 16 (1998.88) & VLBA           & 0.53$\times$0.35, $-$14.9          & 4.0 & x&  & x&              \\
BM112D\dotfill& 1998 December 24           & VLBA           &                                    &           & x& x&  & miscorrelated\\
BM112E\dotfill& 1999 January 28 (1999.08)  & VLBA           & 0.51$\times$0.34, \phantom{1}$-$6.0& 4.4 & x&  & x&              \\
BM112F\dotfill& 1999 March 19 (1999.21)    & VLBA           & 0.48$\times$0.34, $-$10.0          & 4.1 & x& x&  &              \\
BM112G\dotfill& 1999 May 18 (1999.38)      & VLBA           & 0.50$\times$0.35, $-$10.8          & 4.8 & x&  & x&              \\
BM112H\dotfill& 1999 May 26 (1999.40)      & VLBA, VLA, EFLS& 0.43$\times$0.26, $-$31.3          & 2.3 & x& x& x&              \\
BM112I\dotfill& 1999 July 15               & VLBA           &                                    &           & x& x&  & corrupted    \\
BM112J\dotfill& 1999 September 15 (1999.71)& VLBA           & 0.51$\times$0.32, \phantom{1}$-$5.4& 5.8 & x&  & x&              \\
BM112K\dotfill& 1999 October 29 (1999.83)  & VLBA           & 0.50$\times$0.34, $-$19.2          & 4.3 & x& x&  &              \\
BM112L\dotfill& 2000 January 07 (2000.02)  & VLBA           & 0.50$\times$0.34, $-$12.0          & 3.6 & x&  & x&              \\
BM112M\dotfill& 2000 January 30 (2000.08)  & VLBA           & 1.25$\times$0.35, \phantom{$+$}38.8& 4.3 & x& x&  &              \\
BM112N\dotfill& 2000 March 04 (2000.17)    & VLBA           & 0.48$\times$0.33, $-$10.9          & 3.9 & x&  & x&              \\
BM112O\dotfill& 2000 April 12 (2000.28)    & VLBA           & 0.63$\times$0.37, $-$24.4          & 5.0 & x& x&  &              \\
BM112P\dotfill& 2000 May 04 (2000.34)      & VLBA           & 0.50$\times$0.34, $-$11.4          & 4.6 & x&  & x&              \\
BG107\dotfill& 2000 August 12 (2000.61)    & VLBA, VLA, EFLS& 0.46$\times$0.36, $-$14.3          & 4.7 & x& x& x&              \\
\enddata


\tablenotetext{a}{VLBA: Very Long Baseline Array; VLA: 27$\times$25-m Very 
Large Array in Socorro, NM; EFLS: Max-Planck-Institut f\"ur Radioastronomie 
100-m antenna in Effelesberg, Germany
}
\tablenotetext{b}{Average restoring beam full-width half-power and position 
angle, measured east of north.  The unusually large synthesized beam in BM112M 
is due to the loss of two antennas, HN and NL.  See data tables for more detail.
}
\tablenotetext{c}{Average rms noise (scaled to a 1\,km\,s$^{-1}$ channel spacing) 
in emission-free portions of synthesis images.  See Table 5 for more detail. }

\tablenotetext{d}{Velocity range observed. S: low-velocity emission; R: red-shifted 
emission; B: blue-shifted emission
}

\end{deluxetable}

\clearpage

\begin{deluxetable}{ccccrc}
\tabletypesize{\scriptsize}
\tablewidth{0pt}
\tablecaption{Instrument Setup}
\tablehead{
\colhead{Experiment Code\tablenotemark{a}}             & 
\colhead{Bandwidth\tablenotemark{b}}                   &
\colhead{Emission Type}                                & 
\colhead{Bandpass}                                     &
\colhead{Band center v$_{\rm LSR}$\tablenotemark{c}} & 
\colhead{Tuning of Bandpass}\\
&
\colhead{(MHz)}&
&
&
\colhead{(km s$^{-1}$)}&
\\          
}

\startdata
BM056C; BM081A,B; BM112A,H& \phantom{1}8& low-velocity& \phantom{1}1&    618.89\ \ \ \ \ \ \ \ \ & LCP\\
                          &             &             & \phantom{1}2&    533.95\ \ \ \ \ \ \ \ \ & LCP\\
                          &             &             & \phantom{1}3&    511.24\ \ \ \ \ \ \ \ \ & LCP\\
                          &             &             & \phantom{1}4&    426.30\ \ \ \ \ \ \ \ \ & LCP\\
                          &             & red-shifted & \phantom{1}5&   1447.41\ \ \ \ \ \ \ \ \ & RCP\\
                          &             &             & \phantom{1}6&   1359.37\ \ \ \ \ \ \ \ \ & RCP\\
                          &             &             & \phantom{1}7&   1339.76\ \ \ \ \ \ \ \ \ & RCP\\
                          &             &             & \phantom{1}8&   1251.72\ \ \ \ \ \ \ \ \ & RCP\\
                          &             & blue-shifted& \phantom{1}9& $-$336.09\ \ \ \ \ \ \ \ \ & RCP\\
                          &             &             &           10& $-$424.00\ \ \ \ \ \ \ \ \ & RCP\\
                          &             &             &           11& $-$443.74\ \ \ \ \ \ \ \ \ & RCP\\
                          &             &             &           12& $-$531.65\ \ \ \ \ \ \ \ \ & RCP\\
\tableline
BG107                     & \phantom{1}8& low-velocity& \phantom{1}1&    665.00\ \ \ \ \ \ \ \ \ & LCP\\
                          &             &             & \phantom{1}2&    580.00\ \ \ \ \ \ \ \ \ & LCP\\
                          &             &             & \phantom{1}3&    495.00\ \ \ \ \ \ \ \ \ & LCP\\
                          &             &             & \phantom{1}4&    410.00\ \ \ \ \ \ \ \ \ & LCP\\
                          &             & red-shifted & \phantom{1}5&   1485.00\ \ \ \ \ \ \ \ \ & RCP\\
                          &             &             & \phantom{1}6&   1400.00\ \ \ \ \ \ \ \ \ & RCP\\
                          &             &             & \phantom{1}7&   1315.00\ \ \ \ \ \ \ \ \ & RCP\\
                          &             &             & \phantom{1}8&   1230.00\ \ \ \ \ \ \ \ \ & RCP\\
                          &             & blue-shifted& \phantom{1}9& $-$255.00\ \ \ \ \ \ \ \ \ & RCP\\
                          &             &             &           10& $-$340.00\ \ \ \ \ \ \ \ \ & RCP\\
                          &             &             &           11& $-$425.00\ \ \ \ \ \ \ \ \ & RCP\\
                          &             &             &           12& $-$510.00\ \ \ \ \ \ \ \ \ & RCP\\
\tableline
BM112B,F,K,M,O            &           16& low-velocity& \phantom{1}1&    836.00\ \ \ \ \ \ \ \ \ & LCP\\
                          &             &             & \phantom{1}2&    653.00\ \ \ \ \ \ \ \ \ & LCP\\
                          &             &             & \phantom{1}3&    470.00\ \ \ \ \ \ \ \ \ & RCP\\
                          &             &             & \phantom{1}4&    470.00\ \ \ \ \ \ \ \ \ & LCP\\
                          &             & red-shifted & \phantom{1}5&   1568.00\ \ \ \ \ \ \ \ \ & LCP\\
                          &             &             & \phantom{1}6&   1385.00\ \ \ \ \ \ \ \ \ & LCP\\
                          &             &             & \phantom{1}7&   1202.00\ \ \ \ \ \ \ \ \ & LCP\\
                          &             &             & \phantom{1}8&   1019.00\ \ \ \ \ \ \ \ \ & LCP\\
\tableline
BM112C,E,G,J,L,N,P        &           16& low velocity& \phantom{1}1&    500.00\ \ \ \ \ \ \ \ \ & RCP\\
                          &             &             & \phantom{1}2&    500.00\ \ \ \ \ \ \ \ \ & LCP\\
                          &             &             & \phantom{1}3&    317.00\ \ \ \ \ \ \ \ \ & LCP\\
                          &             &             & \phantom{1}4&    134.00\ \ \ \ \ \ \ \ \ & LCP\\
                          &             & blue-shifted& \phantom{1}5&  $-$49.00\ \ \ \ \ \ \ \ \ & LCP\\
                          &             &             & \phantom{1}6& $-$232.00\ \ \ \ \ \ \ \ \ & LCP\\
                          &             &             & \phantom{1}7& $-$415.00\ \ \ \ \ \ \ \ \ & LCP\\
                          &             &             & \phantom{1}8& $-$598.00\ \ \ \ \ \ \ \ \ & LCP\\
\enddata


\tablenotetext{a}{See Table 1 for dates of observation.}
\tablenotetext{b}{The 8 MHz and 16 MHz bands span 108 km s$^{-1}$ and 216 
km s$^{-1}$, respectively.
}
\tablenotetext{c}{Velocities are classical radio velocities in the Local 
Standard of Rest (LSR) reference frame.  
}

\end{deluxetable}

\clearpage

\begin{deluxetable}{ccccccc}
\tabletypesize{\scriptsize}
\tablewidth{0pt}
\tablecaption{Station Positions and Velocities\tablenotemark{a}}
\tablehead{
\colhead{Station}           &
\multicolumn{3}{c}{Station Position} & 
\multicolumn{3}{c}{Error}\\
&
\multicolumn{3}{c}{$\hrulefill$}&
\multicolumn{3}{c}{$\hrulefill$}\\
&
\colhead{X}&
\colhead{Y}&
\colhead{Z}&
\colhead{X}&
\colhead{Y}&
\colhead{Z}\\
&
\colhead{(m)}&
\colhead{(m)}&
\colhead{(m)}&
\colhead{(cm)}&
\colhead{(cm)}&
\colhead{(cm)}
}

\startdata
VLBA-BR\dotfill&         \,$-$2112064.9661&       $-$3705356.5115& 4726813.7932&     0.0476&     0.0885&     0.1069\\
EFLS\dotfill&    \phantom{$-$}4033947.4792& \ \ \ \,\,486990.5244& 4900430.8031&     0.2577&     0.0669&     0.3094\\
VLBA-FD\dotfill&           $-$1324009.1202&       $-$5332181.9713& 3231962.4738&     0.0327&     0.1031&     0.0731\\
VLBA-HN\dotfill& \phantom{$-$}1446375.1259&       $-$4447939.6532& 4322306.1181&     0.0395&     0.1275&     0.1237\\
VLBA-KP\dotfill&           $-$1995678.6199&       $-$5037317.7147& 3357328.1296&     0.0453&     0.1104&     0.0832\\
VLBA-LA\dotfill&           $-$1449752.3520&       $-$4975298.5907& 3709123.9270&     0.0290&     0.0830&     0.0689\\
VLBA-MK\dotfill&           $-$5464074.9591&       $-$2495249.1189& 2148296.8444&     0.1700&     0.1071&     0.1037\\
VLBA-NL\dotfill&      \,\,\,$-$130872.2452&       $-$4762317.1215& 4226851.0426&     0.0217&     0.1067&     0.0953\\
VLBA-OV\dotfill&           $-$2409150.1035&       $-$4478573.2283& 3838617.3949&     0.0546&     0.1027&     0.0919\\
VLBA-PT\dotfill&           $-$1640953.7026&       $-$5014816.0276& 3575411.8817&     0.0321&     0.0851&     0.0692\\
VLBA-SC\dotfill& \phantom{$-$}2607848.5236&       $-$5488069.6773& 1932739.5281&     0.1088&     0.2388&     0.1104\\
VLA\dotfill&               $-$1601185.2077&       $-$5041977.1729& 3554875.7033& \ $\approx 1.0$& \ $\approx 1.0$& \ $\approx 1.0$\\
\multicolumn{7}{c}{$\hrulefill$}\\
&&&&&&\\
&&&&&&\\
Station&
\multicolumn{3}{c}{Station Velocity} & 
\multicolumn{3}{c}{Error}\\ 
&
\multicolumn{3}{c}{$\hrulefill$}&
\multicolumn{3}{c}{$\hrulefill$}\\
& X& Y& Z& X& Y& Z\\
& (cm yr$^{-1}$)& (cm yr$^{-1}$)& (cm yr$^{-1}$)& (cm yr$^{-1}$)& (cm yr$^{-1}$)&(cm yr$^{-1}$)\\
\multicolumn{7}{c}{$\hrulefill$}\\
VLBA-BR\dotfill&           $-$1.2800& \phantom{$-$}0.0370&           $-$1.1200& 0.0160& 0.0280& 0.0304\\
EFLS   \dotfill&           $-$1.5250& \phantom{$-$}1.7100& \phantom{$-$}0.8950& 0.0492& 0.0184& 0.0626\\
VLBA-FD\dotfill&           $-$1.1600&           $-$0.2380&           $-$0.7040& 0.0102& 0.0234& 0.0194\\
VLBA-HN\dotfill&           $-$1.4430&           $-$0.1430& \phantom{$-$}0.0850& 0.0142& 0.0436& 0.0387\\
VLBA-KP\dotfill&           $-$1.2210& \phantom{$-$}0.0150&           $-$0.9740& 0.0149& 0.0309& 0.0255\\
VLBA-LA\dotfill&           $-$1.2850&           $-$0.0170&           $-$0.8300& 0.0083& 0.0174& 0.0165\\
VLBA-MK\dotfill&           $-$1.3600& \phantom{$-$}6.0080& \phantom{$-$}2.9420& 0.0547& 0.0428& 0.0427\\
VLBA-NL\dotfill&           $-$1.4180& \phantom{$-$}0.0090&           $-$0.4710& 0.0095& 0.0289& 0.0245\\
VLBA-OV\dotfill&           $-$1.6270& \phantom{$-$}0.5910&           $-$0.8200& 0.0169& 0.0305& 0.0268\\
VLBA-PT\dotfill&           $-$1.3610&           $-$0.2480&           $-$0.8590& 0.0073& 0.0147& 0.0151\\
VLBA-SC\dotfill& \phantom{$-$}0.9840& \phantom{$-$}0.4390& \phantom{$-$}1.0200& 0.0331& 0.0784& 0.0445\\
VLA\dotfill&               $-$1.3140&           $-$0.1100&           $-$0.8560& \ $\approx 0.1$& $\ \, \approx 0.1$& \ $\approx 0.1$\\
\enddata

\tablenotetext{a}{The VLBA and EFLS station positions and velocities are 
from the United States Naval Observatory's (USNO's) VLBI {\it n9810} 
solution, which has a reference date of January 1, 1997.  USNO's global VLBI 
solutions determine the International Terrestrial Reference Frame (ITRF), the 
International Celestial Reference Frame (ICRF), and Earth Orientation 
Parameters (EOPs).  These solutions, therefore, yield a consistent set of 
station positions and velocities, source positions, and Earth Orientation 
Parameters.  The VLA station position and velocity is from an NRAO 12/00 
geodesy solution (Craig Walker, private communication).  It has been rotated 
to the ITRF reference frame and also has a reference date of 1997 January 1.
}

\end{deluxetable}

\clearpage

\begin{deluxetable}{lccl}
\tabletypesize{\scriptsize}
\rotate
\tablewidth{0pt}
\tablecaption{Largest Sources of Relative Position Error in Synthesis Maps\tablenotemark{a}}
\tablehead{
\colhead{\ \ Type of Error}                         & 
\colhead{\ \ Magnitude over 75 MHz\tablenotemark{b}}&
\colhead{\ \ Magnitude over 4 MHz\tablenotemark{c}} & 
\colhead{\ \ Comments}\\
&                  
\colhead{($\mu$as)}&
\colhead{($\mu$as)}& 
\\
}

\startdata
{\bf Random}\tablenotemark{d}                                      & 1                     & 1                    & 1 Jy peak (strong)\\  
                                                                   & 25\phantom{5}         & 25\phantom{5}        & 50 mJy peak (weak)\\
\tableline
                                                                   &                       &                      &\\
{\bf Systematic (full frequency range):}                           &                       &                      &\\
                                                                   &                       &                      &\\
\ \ \ Positional Uncertainty of Reference Feature\tablenotemark{e} & 10\phantom{5}         & $<$ 1\phantom{15}    & Uncertainty in absolute position of reference feature:\\
                                                                   &                       &                      & \ \ \ $\alpha = 0\asec 0003,\,\delta = 0\as 003$\\
\ \ \ Atmospheric Delay Error\tablenotemark{f}                     & 6                     &$<$ 1\phantom{15}&\\
&&&\\
\ \ \ Frequency Standard Uncertainty\tablenotemark{g}              & 2                     &$<$ 1\phantom{15}&\\
&&&\\
\ \ \ Positional Uncertainty of Calibrator\tablenotemark{h}        & 1                     &$<$ 1\phantom{15}     & Uncertainty in absolute position of calibrators:\\
                                                                   &                       &                      & \ \ \ $\alpha = 0\asec 000042,\,\delta = 0\as 00047$\ (\,4C39.25\,),\\
                                                                   &                       &                      & \ \ \ $\alpha = 0\asec 000114,\,\delta = 0\as 00027$\ (1150$+$81),\\
                                                                   &                       &                      & \ \ \ $\alpha = 0\asec 000020,\,\delta = 0\as 00026$\ (1308$+$33)\\
\ \ \ Baseline Coordinate Uncertainty\tablenotemark{i}             & 1                     &$<$ 1\phantom{15}&\\
\tableline
                                                                   &                       &                        &\\
{\bf Systematic (relative frequencies):}                           &                       &                        &\\
                                                                   &                       &                        &\\
\ \ \ Clock Uncertainty\tablenotemark{j}                           &\ldots                 & 2&\\ 

\enddata

\tablenotetext{a}{The errors discussed in this table are analytic and affect all epochs.  A few epochs 
are subject to an additional, non-analytic source of systematic error, which is discussed in 
$\S 3.2$ of the text and in Table 7.
}

\tablenotetext{b}{75 MHz is the characteristic frequency separation between 
low-velocity and red-shifted or low-velocity and blue-shifted masers.  Since everything 
is referenced to a low-velocity maser, this is the maximum frequency separation 
that need be considered.
}

\tablenotetext{c}{The full 75 MHz frequency span is applicable to single epoch 
position offsets only.  When comparing one epoch's position offsets to 
another's (as with proper motions), a 4 MHz frequency span is more realistic.  
This is because the reference channels for the epochs in Table 1 span only 4 
MHz, meaning that systematic errors largely cancel for red- and blue-shifted 
masers.
}

\tablenotetext{d}{The formal uncertainty in the fitted position of an 
unblended maser spot in a single channel is given by $\Delta\theta_{\rm fit} 
\approx 0.5 \theta_{\rm b} (\Delta S/S$), where $\theta_{\rm b}$ is the 
synthesized beam size, $\Delta S$ is the image rms, and $S$ is the peak 
intensity \citep{rsm88}.  The table values assume $\theta_{\rm b}$ = 500 
$\mu$as and $\Delta S$ = 5 mJy.  Random errors are not frequency dependent 
and do not cancel when comparing position offsets of a given maser feature 
between epochs.
}

\tablenotetext{e}{The relative position error in synthesis maps due to the 
uncertainty in the absolute position of the reference feature is given by 
($\Delta\nu/\nu)\Delta\vec \theta_{\rm ref}$, where $\Delta\nu = \nu - 
\nu_{\rm ref}$, $\nu_{\rm ref}$ is the reference frequency, and 
$\Delta\vec \theta_{\rm ref}$ is the uncertainty in the absolute position of 
the reference feature \citep{tms01}.
}

\tablenotetext{f}{The relative position error in synthesis maps due to the 
atmospheric delay error remaining {\it after delay calibration corrections 
have been applied to the maser source} is given by $(c/B)\ (\Delta\nu/\nu)\ 
\delta\tau_A$, where $c$ is the speed of light, $B$ is the baseline length, 
and $\delta\tau_A$ is the atmospheric delay error \citep{tms01}.  Other 
parameters are defined above.  $\delta\tau_A(nsec) \approx sec\,z\ tan\,z\ 
\Delta z$, where $z$ is the zenith angle and $\Delta z$ is the difference in 
zenith angle between the delay calibrator used to make the corrections and the 
maser source.  Taking average maser-calibrator offsets and zenith angles of 
$\approx 15^{\circ}$ and $\approx 35^{\circ}$, respectively, we get a 6 $\mu 
as$ relative position error for features 75 MHz from the reference feature on 
an 8000 km baseline, appropriate for EFLS$-$VLA, the most sensitive baseline.  
}

\tablenotetext{g}{In general, errors in station clocks consist of two 
components: a fixed offset and a time dependent term arising from frequency 
standard errors.  Hydrogen maser frequency standards typically have a 
fractional uncertainty of 10$^{-14}$, which leads to a residual fringe rate of 
1 mHz.  \citet{tms01} show that the resulting relative 
position error in synthesis maps is given by $(c/B)\ (\Delta\nu/\nu)\ 
(\delta(\delta\nu)/\nu_{\rm 0})\ t$, where $\delta(\delta\nu)/\nu_{\rm 0}$ is 
the fractional fringe rate uncertainty with $\nu_0$ being the LO frequency, 
and $t$ is the time between calibrator scans.  Other parameters are defined 
above.  In computing table values, we assumed a 20-minute calibrator 
separation and an 8000 km baseline.
}

\tablenotetext{h}{An uncertainty in absolute calibrator position leads to a 
residual fringe rate (in mHz) of 0.13\ ($B/5\times 10^3$ km)\ 
($\Delta\theta_{cal}$/1 mas), where $B$ is the baseline length and 
$\Delta\theta_{cal}$ is the position error of the calibrator \citep{tms01}. 
When applied to the line source, \citet{tms01} show that this leads to 
relative position errors in synthesis maps with the same functional dependence 
as above, but with $t$ being the total integration time rather than the time 
between calibrator scans.  It is difficult to assess the combined error of the 
multiple calibrators and many baselines in synthesis maps, but it is likely 
that the dominant contribution in most high sensitivity epochs comes from an 
approximately 2.5 hour 4C39.25 time span on the EFLS$-$VLA baseline, which is 
the most sensitive.  We note that calibrator coordinate errors are given with 
respect to the International Celestial Reference Frame (ICRF) and are 
consistent with the Earth Orientation Parameters (EOPs) used during 
correlation and subsequent data analysis.
}

\tablenotetext{i}{The relative position error in synthesis maps due to a 
baseline error is given by ($\Delta\nu/\nu$) ($\Delta B/B$), where 
$\Delta\nu$ is defined as above, $\Delta B$ is the baseline error, and $B$ is 
the baseline length \citep{tms01}.  The EFLS$-$VLA baseline is representative, 
since, being the most sensitive, it is expected to have the largest effect on 
synthesis maps.  As mentioned in the text, the position of the VLA is 
uncertain to 3 cm, while all other station positions are uncertain to 1 mm.  
Baseline coordinates are consistent with the ICRF and the EOPs (see note 
above).
}

\tablenotetext{j}{Clock offset errors do not scale with the full frequency 
range, but rather with the relative frequency of remote and reference channels, 
each with respect to their own LO frequencies \citep{tms01}.  Hence, 
$\Delta\nu_{rel} = (\nu - \nu_0) - (\nu_{ref} - \nu_{ref_0})$.  The relative 
position error in synthesis maps due to clock errors is then 
$(\Delta\nu_{rel}/\nu) (c \delta\tau_0/B)$, where $\delta\tau_0$ is the 
post calibration delay residual.  The accuracy of delay measurements is 
expected to be of order $(1/2\Delta\nu_{bw})$ ({\it 1/SNR}), where 
$\Delta\nu_{bw}$ is the total bandwidth and {\it SNR} is the signal-to-noise 
ratio of calibrator measurements \citep{tms01}.  Assuming {\it SNR} = 150, we 
obtain a residual delay of 0.4 nsec and a relative position error in synthesis 
maps of 2 $\mu$as for relative frequency separations of 4 MHz.
}
\end{deluxetable}

\clearpage

\begin{deluxetable}{crrrrrrrrrrrrrrr}
\tabletypesize{\scriptsize}
\rotate
\tablewidth{0pt}
\tablecaption{Channel Data}
\tablehead{
\colhead{channel}& 
\colhead{$v_{\rm LSR}$\tablenotemark{a}}&
\colhead{$v_{\rm rel}$\tablenotemark{a}}&
\colhead{F$_{\nu}$\tablenotemark{b}}&
\colhead{$\sigma_{{\rm F}_{\nu}}$\tablenotemark{b}}&
\colhead{$\Delta x$\tablenotemark{c}}&
\colhead{$\sigma_x$\tablenotemark{c}}&
\colhead{$\Delta y$\tablenotemark{c}}&
\colhead{$\sigma_y$\tablenotemark{c}}&
\colhead{maj\tablenotemark{d}}&
\colhead{$\sigma_{\rm maj}$\tablenotemark{d}}&
\colhead{min\tablenotemark{d}}&
\colhead{$\sigma_{\rm min}$\tablenotemark{d}}&
\colhead{pa\tablenotemark{d}}&
\colhead{$\sigma_{\rm pa}$\tablenotemark{d}}&
\colhead{$v_{\rm band}$\tablenotemark{e}}\\

\colhead{(1$-$512)}& 
\colhead{(km s$^{-1}$)}&
\colhead{(km s$^{-1}$)}&
\colhead{(Jy)}&
\colhead{(mJy)}&
\colhead{(mas)}&
\colhead{(mas)}&
\colhead{(mas)}&
\colhead{(mas)}&
\colhead{(mas)}&
\colhead{(mas)}&
\colhead{(mas)}&
\colhead{(mas)}&
\colhead{($^{\circ}$)}&
\colhead{($^{\circ}$)}&
\colhead{(km s$^{-1}$)}\\
}

\startdata

177&  1464.157&  1467.702&  0.038&     3.690&   4.458&   0.024&   0.784&   0.032&   0.82&   0.08&   0.57&   0.05&  167&  10&  1447.41\\
178&  1463.946&  1467.490&  0.054&     3.850&   4.486&   0.018&   0.783&   0.023&   0.71&   0.05&   0.53&   0.04&    7&  10&  1447.41\\
179&  1463.735&  1467.279&  0.057&     3.770&   4.499&   0.017&   0.750&   0.022&   0.71&   0.05&   0.54&   0.04&   19&   9&  1447.41\\
180&  1463.525&  1467.067&  0.044&     3.680&   4.508&   0.021&   0.753&   0.028&   0.68&   0.06&   0.60&   0.05&   23&  27&  1447.41\\
223&  1454.466&  1457.964&  0.027&     3.570&   4.475&   0.033&   0.778&   0.043&   0.71&   0.09&   0.42&   0.06&   12&  10&  1447.41\\
224&  1454.255&  1457.752&  0.039&     3.550&   4.503&   0.023&   0.791&   0.030&   0.65&   0.06&   0.49&   0.04&   14&  13&  1447.41\\
225&  1454.045&  1457.541&  0.057&     3.560&   4.531&   0.016&   0.776&   0.020&   0.65&   0.04&   0.55&   0.03&   14&  15&  1447.41\\
226&  1453.834&  1457.329&  0.079&     3.620&   4.523&   0.012&   0.782&   0.015&   0.68&   0.03&   0.55&   0.03&   24&   8&  1447.41\\
227&  1453.623&  1457.117&  0.092&     3.730&   4.513&   0.010&   0.800&   0.013&   0.71&   0.03&   0.55&   0.02&   15&   6&  1447.41\\
228&  1453.413&  1456.906&  0.102&     3.720&   4.512&   0.009&   0.814&   0.012&   0.69&   0.03&   0.55&   0.02&    9&   6&  1447.41\\
229&  1453.202&  1456.694&  0.106&     3.620&   4.515&   0.009&   0.802&   0.011&   0.68&   0.02&   0.53&   0.02&    8&   5&  1447.41\\
230&  1452.991&  1456.482&  0.112&     3.610&   4.517&   0.008&   0.803&   0.011&   0.65&   0.02&   0.52&   0.02&    8&   6&  1447.41\\
231&  1452.781&  1456.271&  0.114&     3.560&   4.515&   0.008&   0.818&   0.010&   0.65&   0.02&   0.54&   0.02&    6&   6&  1447.41\\
232&  1452.570&  1456.059&  0.115&     3.680&   4.538&   0.008&   0.823&   0.011&   0.66&   0.02&   0.55&   0.02&   17&   7&  1447.41\\
233&  1452.359&  1455.847&  0.116&     3.680&   4.546&   0.008&   0.821&   0.010&   0.69&   0.02&   0.52&   0.02&    9&   4&  1447.41\\

\enddata



\tablecomments{Table 5 provides the Channel data for all the observations and
is published in its entirety in the 
electronic edition of the {\it Astrophysical Journal}. A portion is shown 
here for guidance regarding its form and content.}

\tablenotetext{a}{Velocities according to the classical and relativistic 
definitions of the Doppler shift in the LSR frame.  See $\S$2.6 for a detailed 
discussion of conversion from classical to relativistic LSR velocity.  
Classical Doppler tracking was implemented using the AIPS task CVEL, which is 
accurate to $< 0.004$\,km\,s$^{-1}$ \citep{bgm00}.}

\tablenotetext{b}{Fitted peak intensity (Jy beam$^{-1}$) is reported in this table.  
Note that Flux density (Jy) = Fitted peak intensity (Jy beam$^{-1}$) $\times 
( \theta^2_{meas}/\theta^2_{beam} )$, where $\theta^2_{meas}$ is the measured maser spot 
area and $\theta^2_{beam}$ is the area of the restoring beam.  Since maser spots are assumed 
to be unresolved, these values can be quoted in Jy.}

\tablenotetext{c}{East-west and north-south position offsets and uncertainties, 
measured with respect to the emission at 510.0 km s$^{-1}$ (relativistic).  
Position uncertainties reflect fitted random error only.  The true 
random error is estimated to be the maximum of the fitted random error and the random 
error computed according to Table 4, footnote {\it d}.
}

\tablenotetext{d}{Major axes, minor axes, and position angles of maser spots 
and uncertainties.
}

\tablenotetext{e}{Band identifier refers to the nonrelativistic velocity in 
channel 256.5.  
}
\end{deluxetable}

\clearpage

\begin{deluxetable}{lrcccccc}
\tabletypesize{\scriptsize}
\tablewidth{0pt}
\tablecaption{Continuum Detections}
\tablehead{
\colhead{Epoch\tablenotemark{a}\,}& 
\colhead{Freq.\tablenotemark{b}}&
\colhead{F$_{\nu}$\tablenotemark{c}}&
\colhead{rms\tablenotemark{d}}&
\colhead{$\Delta x$\tablenotemark{e}}&
\colhead{$\sigma_x$\tablenotemark{e}}&
\colhead{$\Delta y$\tablenotemark{e}}&
\colhead{$\sigma_y$\tablenotemark{e}}\\
& 
&
\colhead{(mJy)}&
\colhead{(mJy)}&
\colhead{(mas)}&
\colhead{(mas)}&
\colhead{(mas)}&
\colhead{(mas)}\\
}

\startdata
BM056C&  blue-shifted& \phantom{$<$\ }1.19& 0.18&  $-$0.240& 0.051& 0.952& 0.055\\
BM081B& \ red-shifted& \phantom{$<$\ }1.66& 0.19&  \phantom{$-$}0.108& 0.022& 0.800& 0.032\\
      &  blue-shifted& \phantom{$<$\ }1.28& 0.19&  $-$0.194& 0.028& 1.018& 0.035\\
BM112B& \ red-shifted& \phantom{$<$\ }1.53& 0.19&  \phantom{$-$}0.007& 0.037& 1.360& 0.099\\
BM112C&  blue-shifted& \phantom{$<$\ }1.67& 0.18&  $-$0.190& 0.028& 1.099& 0.070\\
BM112E&  blue-shifted& \phantom{$<$\ }2.40& 0.20&  $-$0.028& 0.030& 0.952& 0.048\\
BM112F& \ red-shifted& \phantom{$<$\ }1.90& 0.19&  $-$0.039& 0.054& 0.897& 0.093\\
BM112G&  blue-shifted& \phantom{$<$\ }0.72& 0.21&  \phantom{$-$}0.200& 0.047& 1.008& 0.080\\
BM112J&  blue-shifted& \phantom{$<$\ }1.70& 0.26&  $-$0.147& 0.075& 1.093& 0.147\\
BM112K& \ red-shifted& \phantom{$<$\ }1.72& 0.19&  $-$0.192& 0.134& 1.534& 0.213\\
BM112L&  blue-shifted& \phantom{$<$\ }1.75& 0.16&  $-$0.009& 0.029& 1.252& 0.061\\
BM112M& \ red-shifted& \phantom{$<$\ }2.68& 0.20&  $-$0.346& 0.363& 0.990& 0.157\\
BM112N&  blue-shifted& \phantom{$<$\ }2.39& 0.18&  $-$0.049& 0.013& 0.871& 0.030\\
BM112O& \ red-shifted& \phantom{$<$\ }4.01& 0.24&  \phantom{$-$}0.022& 0.021& 0.947& 0.040\\
BM112P&  blue-shifted& \phantom{$<$\ }2.52& 0.21&  $-$0.029& 0.061& 1.018& 0.077\\
 BG107& \ red-shifted& \phantom{$<$\ }3.95& 0.22&  $-$0.133& 0.015& 1.050& 0.018\\
      &  blue-shifted& \phantom{$<$\ }3.66& 0.20&  $-$0.152& 0.010& 1.022& 0.013\\

\enddata

\tablenotetext{a}{We report epochs where continuum emission is observed 
at $> 3\sigma$
}

\tablenotetext{b}{For BM056C and BM081B we averaged the inner 490 channels 
of the three independent red-shifted and three independent blue-shifted bands 
(1470 channels).  For all other epochs we averaged the inner 470 channels of 
the four red-shifted and four blue-shifted bands (1960 channels). 
}

\tablenotetext{c}{We report the flux density, since continuum detections 
are partially resolved.  Flux density (mJy) = Fitted peak intensity (mJy beam$^{-1}$) 
$\times ( \theta^2_{meas}/\theta^2_{beam} )$, where $\theta^2_{meas}$ is the measured area 
of the continuum emission and $\theta^2_{beam}$ is the area of the restoring beam.  
Strictly speaking, this is not a total flux density, but rather an angle integrated 
flux density, since we measure it with an interferometer that has a limited range 
of baseline lengths.
}

\tablenotetext{d}{1$\sigma$ noise level in synthesis map.
}

\tablenotetext{e}{East-west and north-south position offsets and uncertainties 
measured with respect to the reference maser feature adopted for the epoch of 
observation.  Position uncertainties reflect random error only.
}

\end{deluxetable}

\clearpage

\begin{deluxetable}{crrrrrrrr}
\tabletypesize{\scriptsize}
\tablewidth{0pt}
\tablecaption{Derived Position Shifts \tablenotemark{a}}
\tablehead{
\colhead{Experiment Code}           &
\multicolumn{4}{c}{Low-velocity Masers} & 
\multicolumn{4}{c}{Red-shifted Masers}\\ 
&
\multicolumn{4}{c}{$\hrulefill$}&
\multicolumn{4}{c}{$\hrulefill$}\\
&
\colhead{x}&
\colhead{$\sigma_x$}&
\colhead{y}&
\colhead{$\sigma_y$}&
\colhead{x}&
\colhead{$\sigma_x$}&
\colhead{y}&
\colhead{$\sigma_y$}\\
&
\colhead{($\mu$as)}&
\colhead{($\mu$as)}&
\colhead{($\mu$as)}&
\colhead{($\mu$as)}&
\colhead{($\mu$as)}&
\colhead{($\mu$as)}&
\colhead{($\mu$as)}&
\colhead{($\mu$as)}\\
}

\startdata
BM056C\dotfill& \phantom{$-$}8& 3& \phantom{$-$}1& 4&  \phantom{$-$}5& 5&            $-$1& 6\\
BM081A\dotfill& \phantom{$-$}5& 1&           $-$3& 2&            $-$8& 4&           $-$10& 6\\
BM081B\dotfill&           $-$4& 2&          $-$11& 2&            $-$1& 3&            $-$3& 5\\
BM112A\dotfill&           $-$8& 2& \phantom{$-$}3& 4& \phantom{$-$}33& 5&  \phantom{$-$}6& 8\\
BM112B\dotfill& \phantom{$-$}5& 3& \phantom{$-$}0& 5& \phantom{$-$} 2& 5&  \phantom{$-$}0& 7\\
BM112C\dotfill& \phantom{$-$}3& 2&           $-$2& 3&                &  &                &   \\
BM112E\dotfill& \phantom{$-$}0& 3& \phantom{$-$}1& 4&                &  &                &      \\
BM112F\dotfill& \phantom{$-$}4& 2&           $-$2& 3&            $-$3& 4&            $-$6& 6\\
BM112G\dotfill&           $-$3& 3& \phantom{$-$}8& 4&                &  &                &      \\
BM112H\dotfill&           $-$6& 1& \phantom{$-$}0& 2&           $-$24& 4&  \phantom{$-$}0& 6\\
BM112J\dotfill&             10& 3&           $-$6& 4&                &  &                &      \\
BM112K\dotfill& \phantom{$-$}1& 2&           $-$1& 3&            $-$5& 6&  \phantom{$-$}3& 8\\
BM112L\dotfill& \phantom{$-$}3& 2&           $-$3& 3&                &  &                &      \\
BM112M\dotfill&           $-$7& 3&           $-$5&10&            $-$5& 4&  \phantom{$-$}0& 14\\
BM112N\dotfill& \phantom{$-$}3& 3& \phantom{$-$}6& 4&                &  &                &      \\
BM112O\dotfill&           $-$7& 4& \phantom{$-$}9& 6&           $-$12& 4& \phantom{$-$}15& 7\\
BM112P\dotfill&           $-$2& 3&          $-$10& 4&                &  &                &      \\
BG107\dotfill&           $-$20& 4& \phantom{$-$}0& 5&           $-$17& 7& \phantom{$-$}37& 9\\
\enddata


\tablenotetext{a}{Position shifts are determined by minimizing the sum of the 
deviations in each epoch from the weighted average of the masers in the medium 
sensitivity epochs.  Weighted averages are computed and deviations determined 
for each frequency channel.  See $\S 3.2$ for details.
}

\end{deluxetable}

\clearpage

\begin{figure}
\epsscale{.80}
\plotone{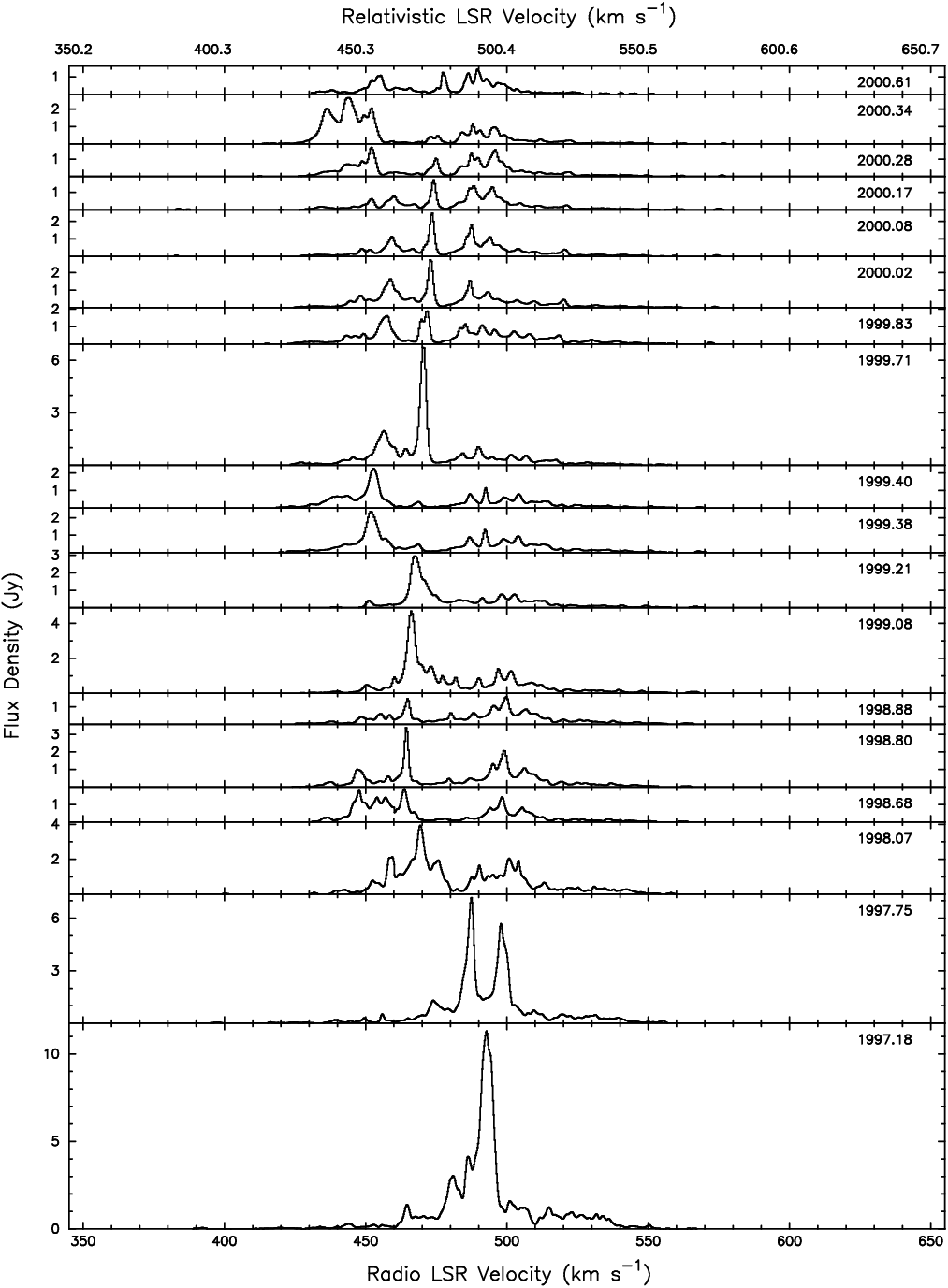}
\caption{Spectra for low-velocity masers.  The vertical scale on the page is
the same for each plot.  The spectra show the total imaged power obtained
from the fitted peak flux densities of maser spots.  Channels
with no detectable emission are not shown. 
Velocities are defined with respect to the Local
Standard of Rest, assuming the nonrelativistic radio definition of Doppler shift (bottom axis) 
and relativistic radio definition of Doppler shift (top axis).
\label{fig1}}
\end{figure}

\clearpage
                                                                     
\begin{figure}
\epsscale{.85}
\plotone{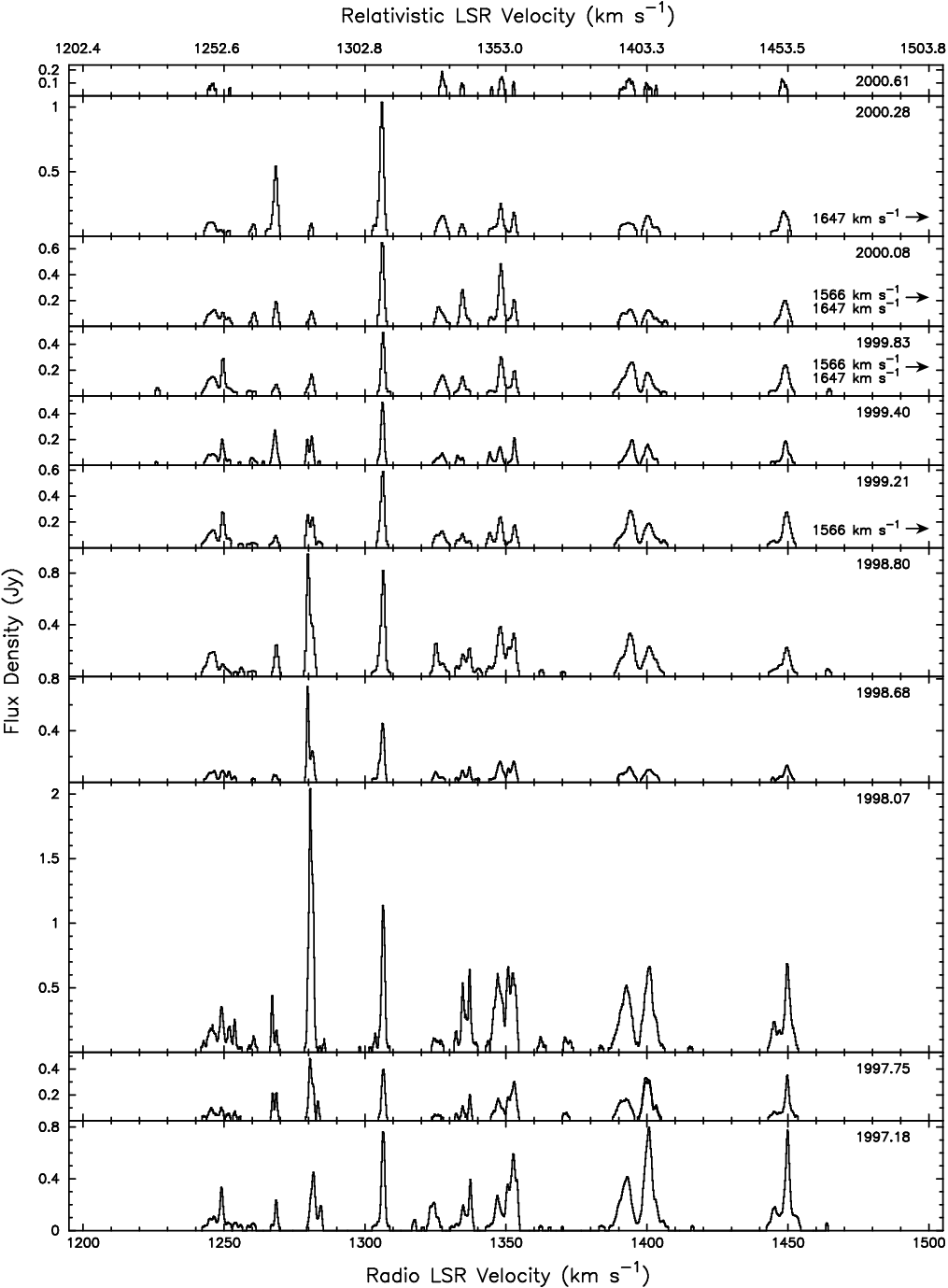}
\caption{Spectra for red-shifted masers, displayed as in Figure~1. The Doppler
components at 1566\,km\,s$^{-1}$ (epochs 1999.21, 1999.83, and 2000.08) and 
1647\,km\,s$^{-1}$ (epochs 1999.83, 2000.08, and 2000.28) are shown separately in 
Figure~3.
\label{fig2}}
\end{figure}

\clearpage
                                                                     
\begin{figure}
\epsscale{.85}
\plotone{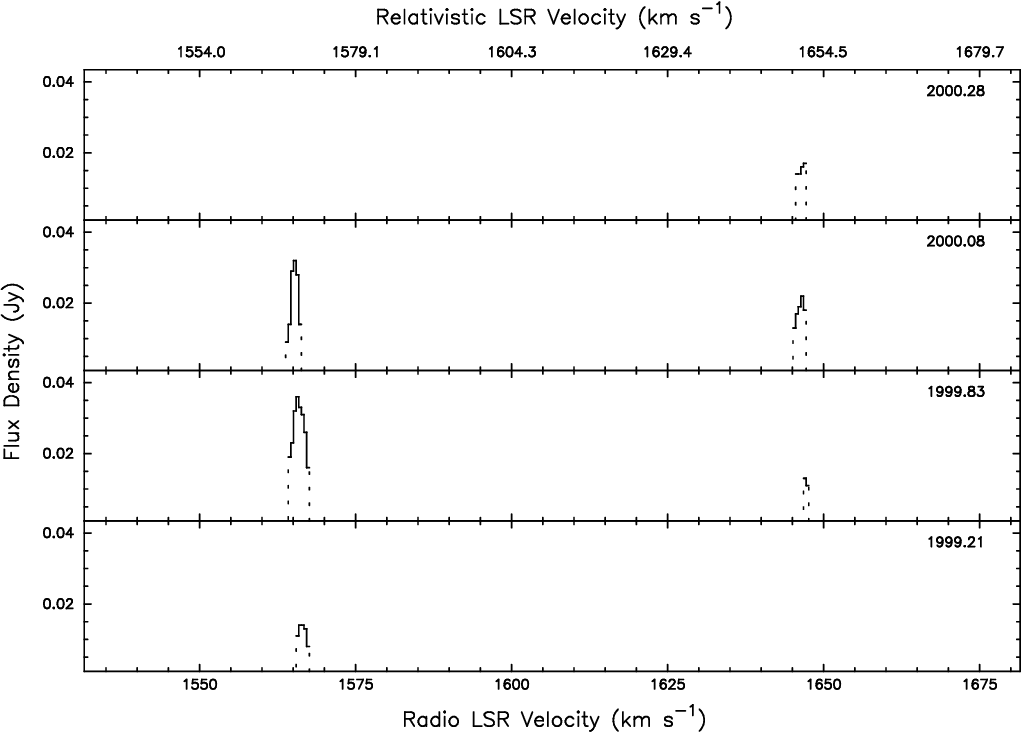}
\caption{Spectra for red-shifted masers at 1566\,km\,s$^{-1}$ and 
1647\,km\,s$^{-1}$, displayed as in Figure~1.  The dotted portions 
of the spectra delineate fits that fall below our $6\sigma$ threshold (see 
text for details).
\label{fig3}}
\end{figure}

\clearpage
                                                                     
\begin{figure}
\epsscale{.85}
\plotone{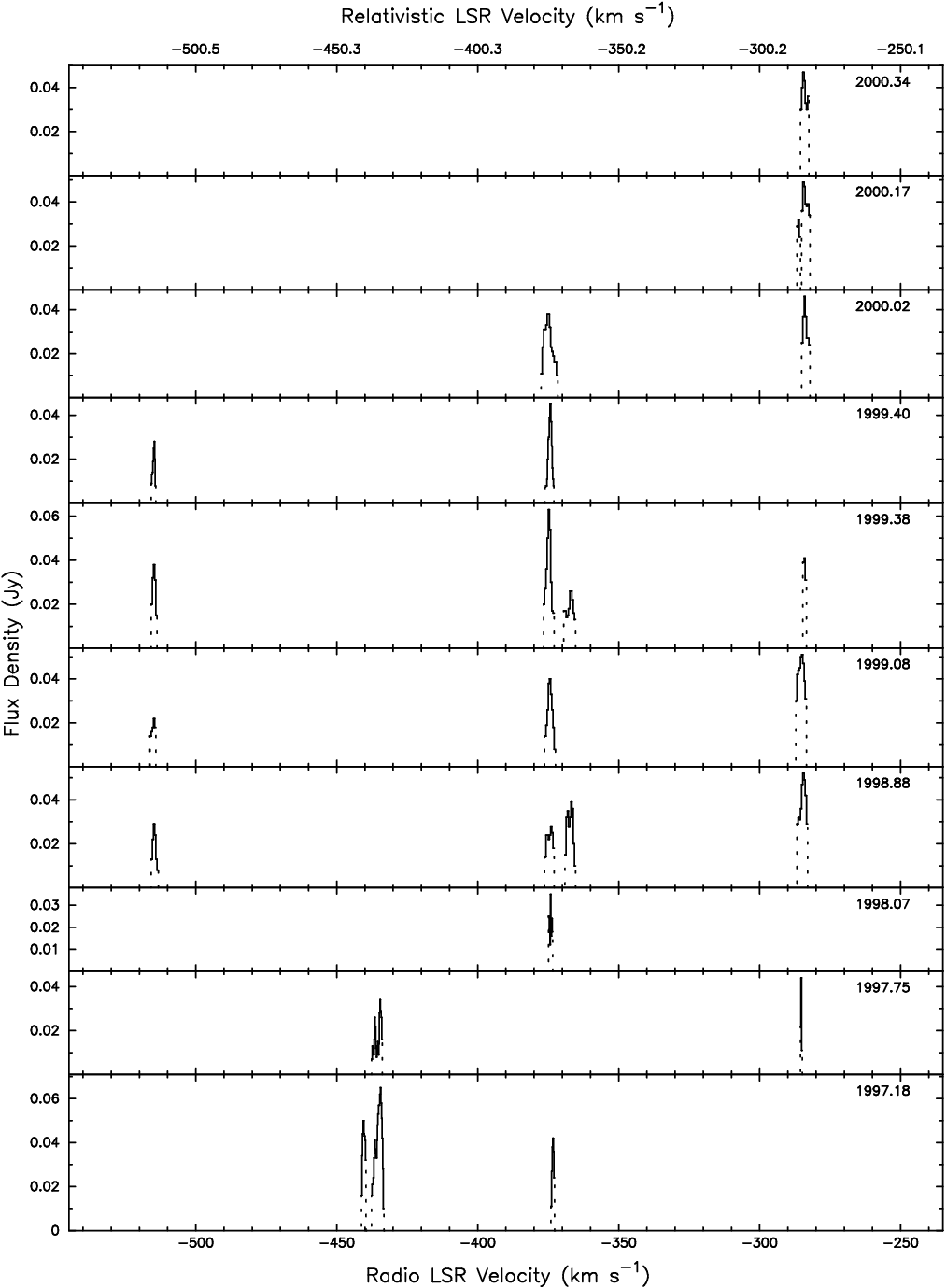}
\caption{Spectra for blue-shifted masers, displayed as in Figure~1. 
The dotted portions of the spectra delineate fits that fall below our $6\sigma$ 
detection threshold (see text for details).\label{fig4}}
\end{figure}

\clearpage
                                                                     
\begin{figure}
\epsscale{.85}
\plotone{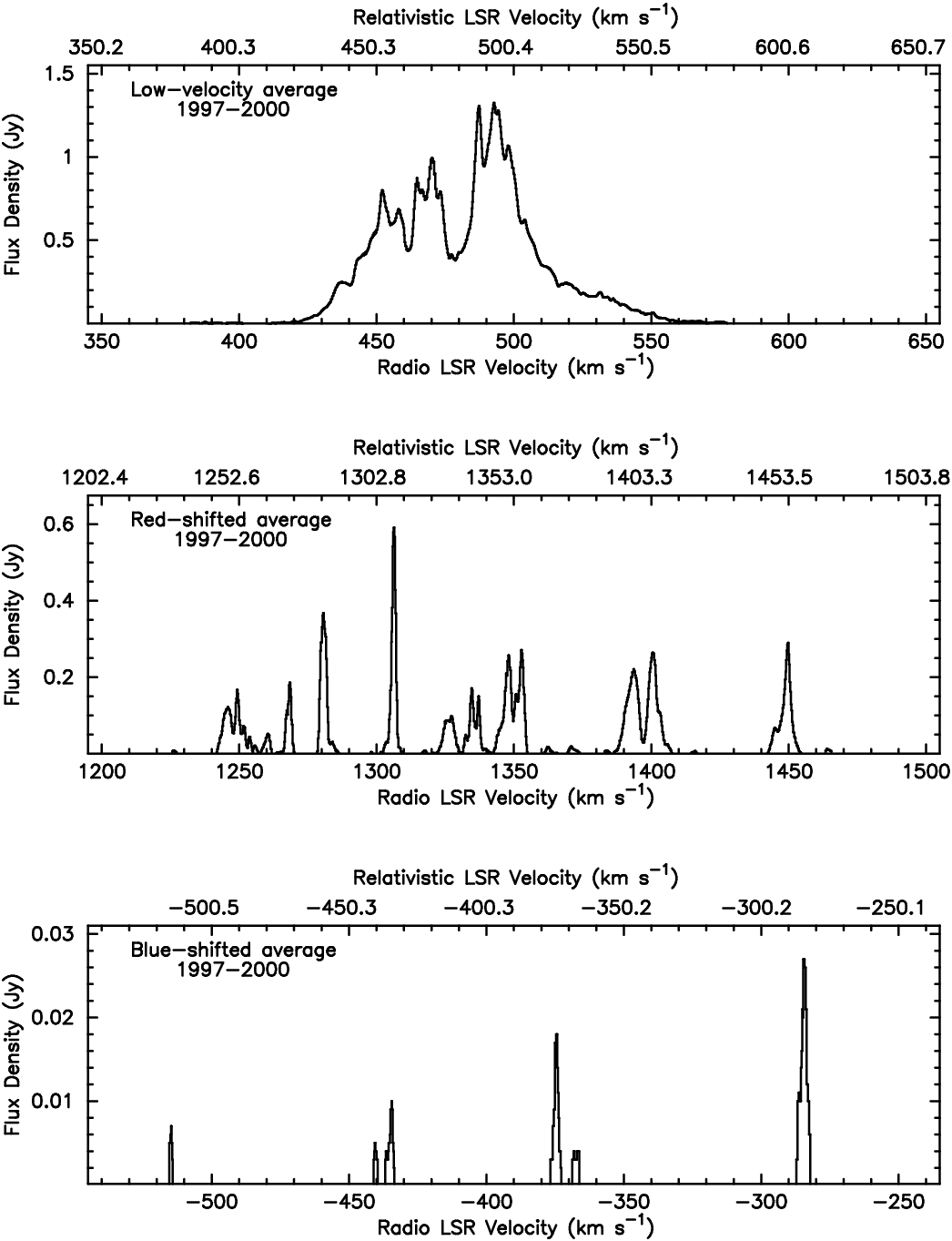}
\caption{Unweighted average maser spectra for all observing epochs (1997-2000),
displayed as in Figure~1. {\it (top)} Low-velocity emission. {\it (middle)}
Red-shifted high-velocity emission. {\it (bottom)} Blue-shifted high-velocity emission. 
LSR velocities are defined assuming the nonrelativistic radio definition of Doppler 
shift (bottom axis) or relativistic radio definition of Doppler shift (top axis). \label{fig5}}
\end{figure}

\clearpage
                                                                     

\clearpage

\begin{figure}
\epsscale{.7}
\plotone{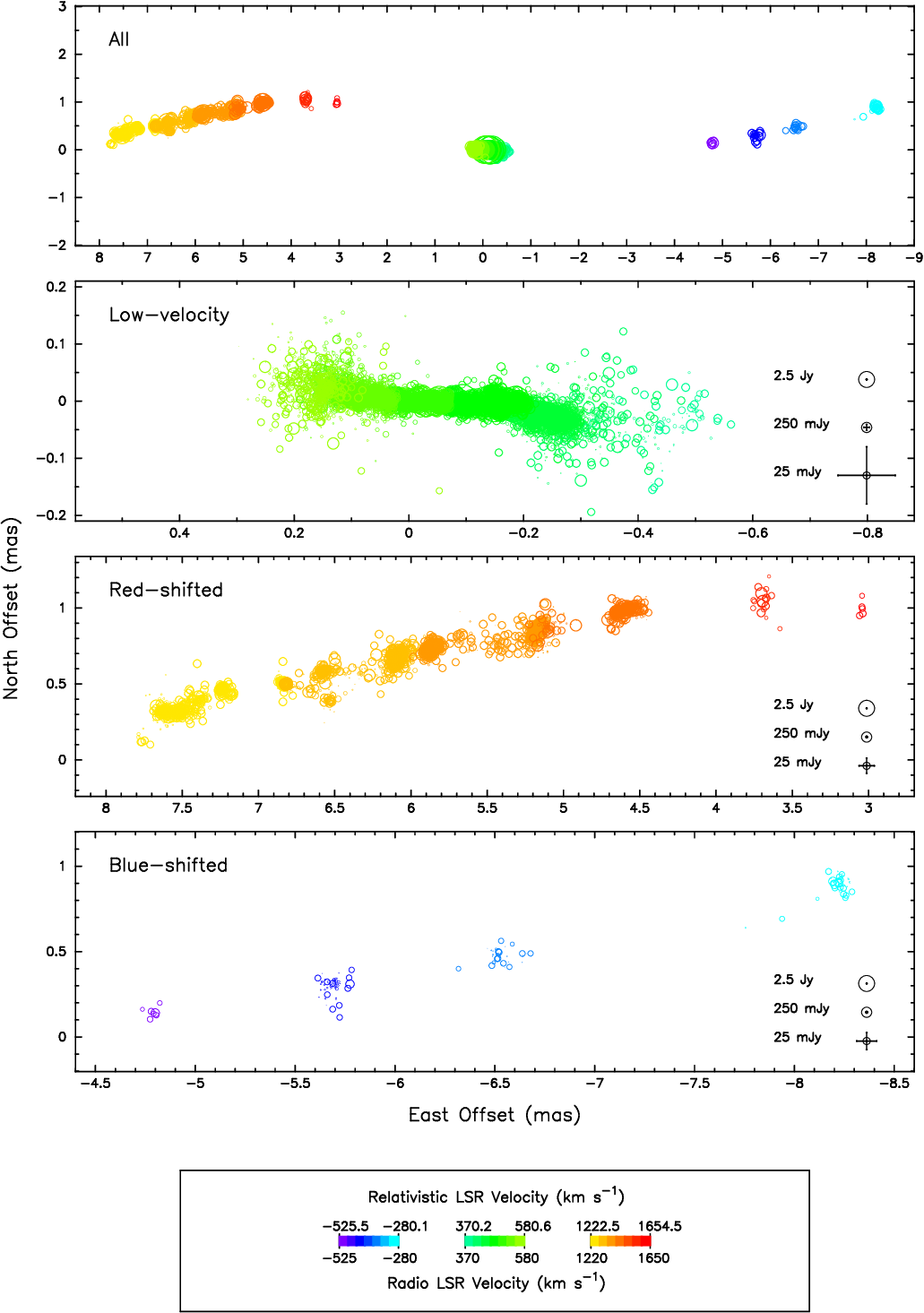}
\caption{Maps of maser emission for all epochs superposed.  {\it (top)} The
full velocity range of the emission.  {\it (upper middle)} Low-velocity
emission.  {\it (lower middle)} Red-shifted high-velocity emission.  {\it
(bottom)} Blue-shifted high-velocity emission.  Symbol diameter is proportional to
the logarithm of peak flux density.  Each panel is scaled separately.  A key on 
the right hand side of the three lower panels gives the symbol sizes of 2.5 Jy, 
250 mJy, and 25 mJy masers, as well as the x-y sizes of their associated random 
errors.  The color bar at bottom codes nonrelativistic LSR velocity (lower label) 
or relativistic LSR velocity (upper label).  Note that the thinness of the maser disk 
in the low-velocity panel is obscured by the large number of masers plotted.  See 
Figures~7-9 for additional discussion.
\label{fig6}} 
\end{figure}

\clearpage

\begin{figure}
\epsscale{.85}
\plotone{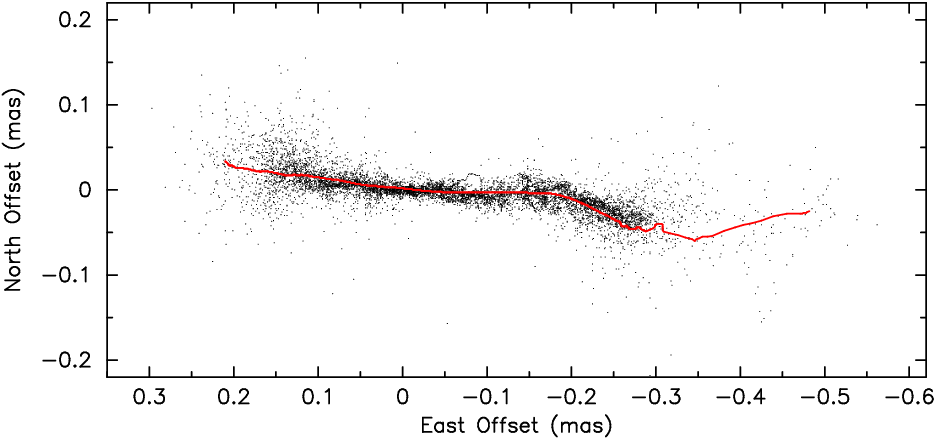}
\caption{Map of low-velocity maser emission.  All masers have the same symbol size.  
The red curve is the boxcar smoothed (30 km s$^{-1}$ width) weighted average 
of low-velocity masers, computed from medium-sensitivity epochs (see $\S 3.2$).  
\label{fig7}} 
\end{figure}

\clearpage

\begin{figure}
\epsscale{1.0}
\plotone{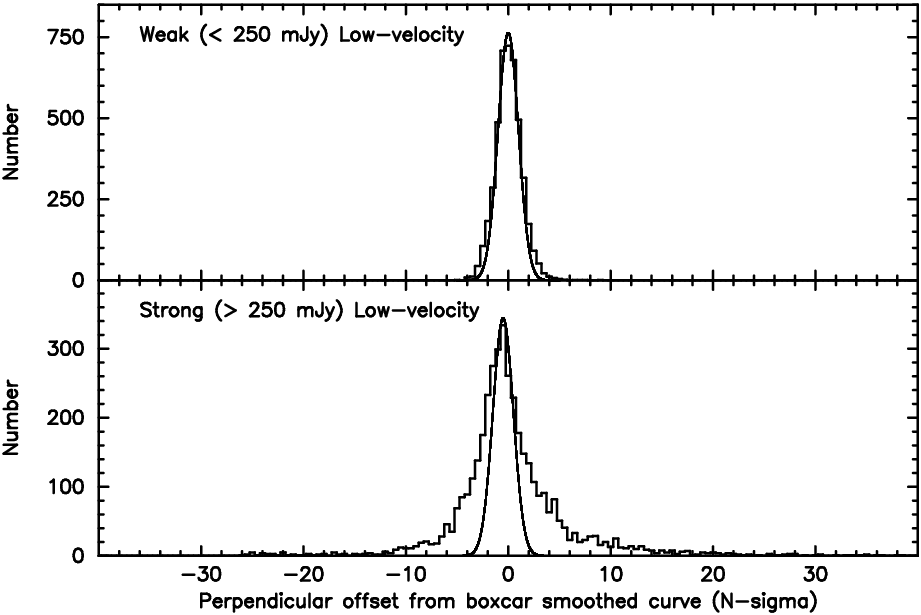}
\caption{Histograms of separations between individual low-velocity maser 
spots and the average emission locus (Figure 7, red curve).  Each offset is 
divided by the corresponding random measurement uncertainty.  All low-velocity 
masers are plotted.  {\it (top)} Low-velocity masers $<250$ mJy, for which 
corresponding measurement uncertainties are $>5\mu$as, plotted.  
The distribution for weak masers is Gaussian (see superposed 
curve), as would be expected from a noise dominated distribution.  
{\it (bottom)} Low-velocity masers $>250$ mJy, for which corresponding 
measurement uncertainties are $<5\mu$as, plotted.  The distribution for strong masers 
exhibits nonGaussian wings and tail (see superposed curve) that may be attributed to 
source structure or systematic error arising from the 
representation of the disk plane position by the average of maser spot 
positions (Figure 7, red curve), at the few $\mu$as level.
\label{fig8}} 
\end{figure}

\clearpage
                                                                    
\begin{figure}
\epsscale{1.0}
\plotone{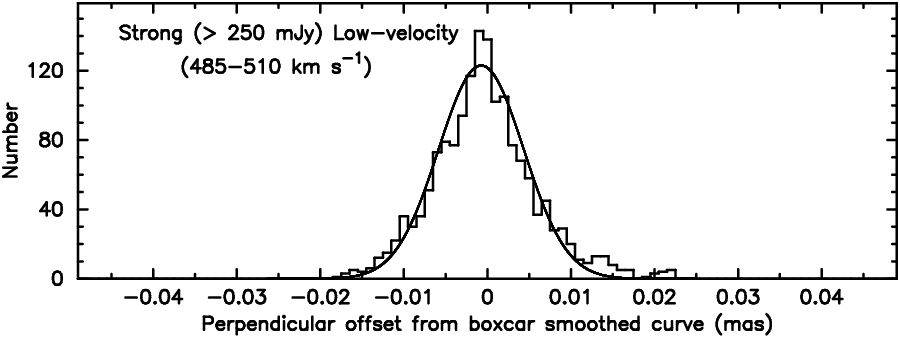}
\caption{Histogram as in Figure 8 for masers $>250$ mJy but for the velocity range 
$485-510$ km s$^{-1}$ (East Offset $-0.1$ to $0.0$ mas).  This is the velocity range 
over which the model accretion disk \citep{hmgt05} is viewed edge-on.  The 
distribution of strong masers is nearly Gaussian (see superposed curve) with 
full-width at half-maximum (FWHM) of $\approx 12 \mu$as ($\sigma \approx 5.1 \mu$as), which we 
interpret to be the underlying accretion disk thickness.  
\label{fig9}} 
\end{figure}

\clearpage
                                                                    
\begin{figure}
\epsscale{.8}
\plotone{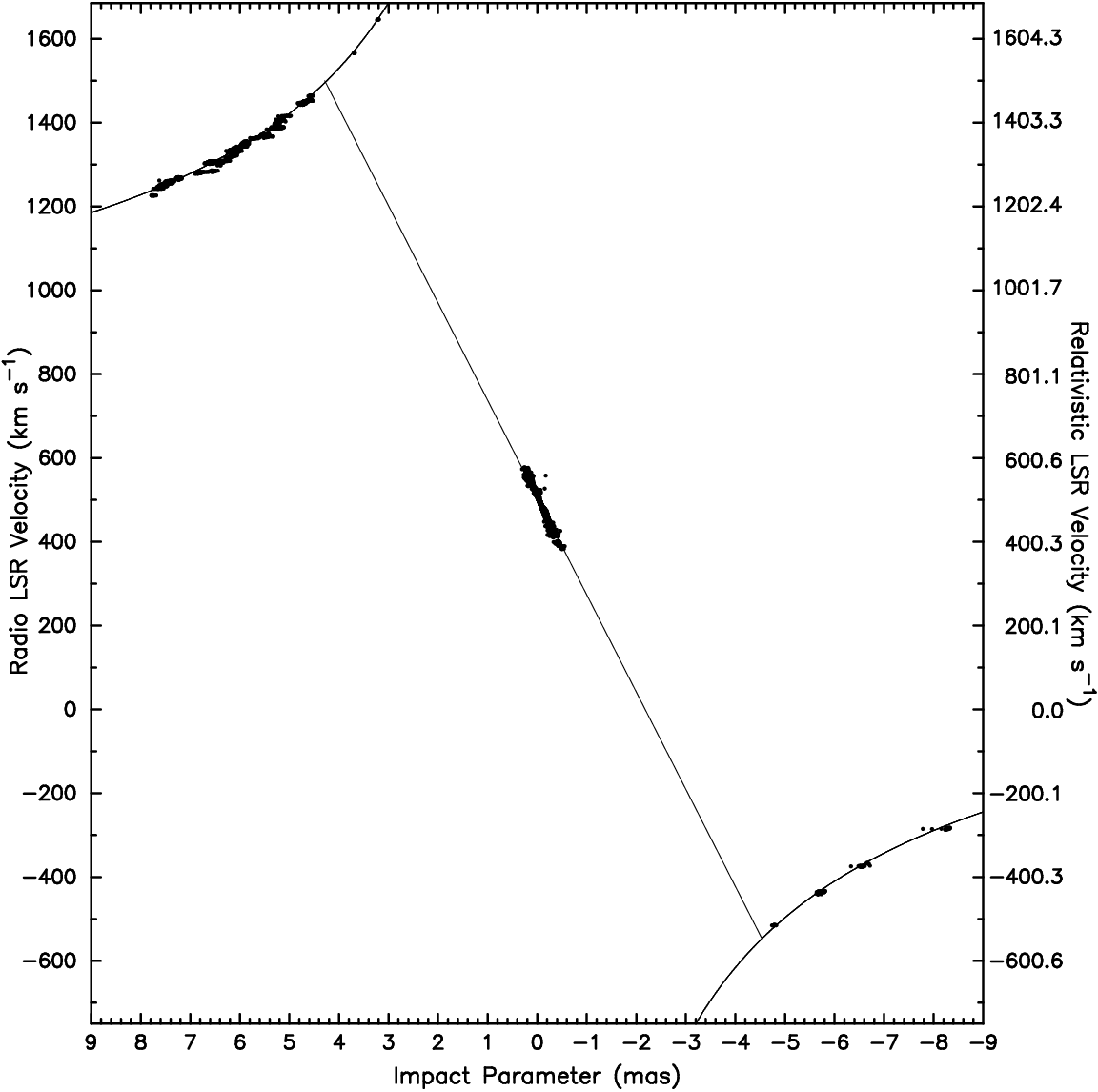}
\caption{Position--velocity diagram for all epochs superposed.  Positive impact
parameters  are to the east.  The linear gradient in low-velocity emission is
characteristic of emission from material in highly inclined circular orbits over a narrow
range of radii. The fitted straight line marks the locus for emission at a single radius.
(A steeper line indicates a smaller radius). High-velocity masers trace declining rotation
curves, which are fitted here in position--velocity space and shown for Keplerian
orbits.  Good agreement with the model $v-v_{sys}\propto |b|^{-0.5}$, where $v_{sys}$ is
systemic velocity and $b$ is impact parameter, suggests that the high-velocity emission
arises close to a single diameter through the underlying disk and that warping is of
second-order importance \citep[e.g.,][]{mmh95}.  We note that the Doppler components at
1566\,km\,s$^{-1}$ and  1647\,km\,s$^{-1}$ appear to arise inside the mean
radius of the low-velocity emission.  Nothing has previously been mapped at these
small radii. LSR velocities are defined assuming the 
nonrelativistic radio definition of Doppler shift (left-hand axis) or relativistic 
radio definition of Doppler shift (right-hand axis). 
\label{fig10}}
\end{figure}

\clearpage
                                                                     
\begin{figure}
\includegraphics[scale=0.85]{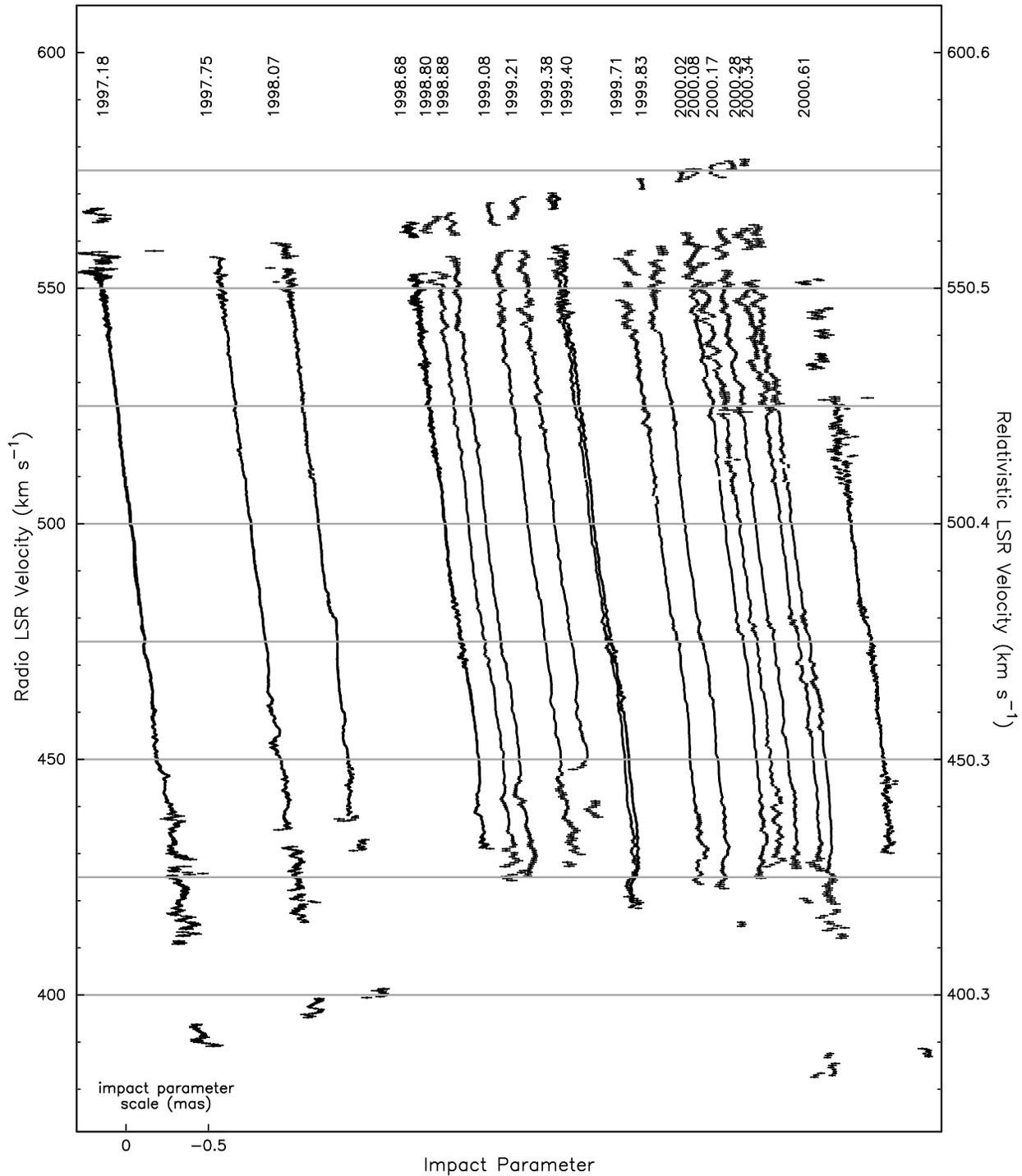}
\caption{Position--velocity traces for low-velocity emission at each epoch
individually.  The impact parameter scale is shown in the bottom left corner of the
figure and the separation between traces is proportional to the time between epochs
(denoted by numeric labels).  Horizontal error bars denote uncertainty in impact
parameter and reflect random error only.  LSR velocities are defined assuming the 
nonrelativistic radio definition 
of Doppler shift (left-hand axis) or relativistic radio definition of Doppler shift 
(right-hand axis).
\label{fig11}}
\end{figure}

\clearpage
                                                                     
\begin{figure}
\includegraphics[scale=0.85]{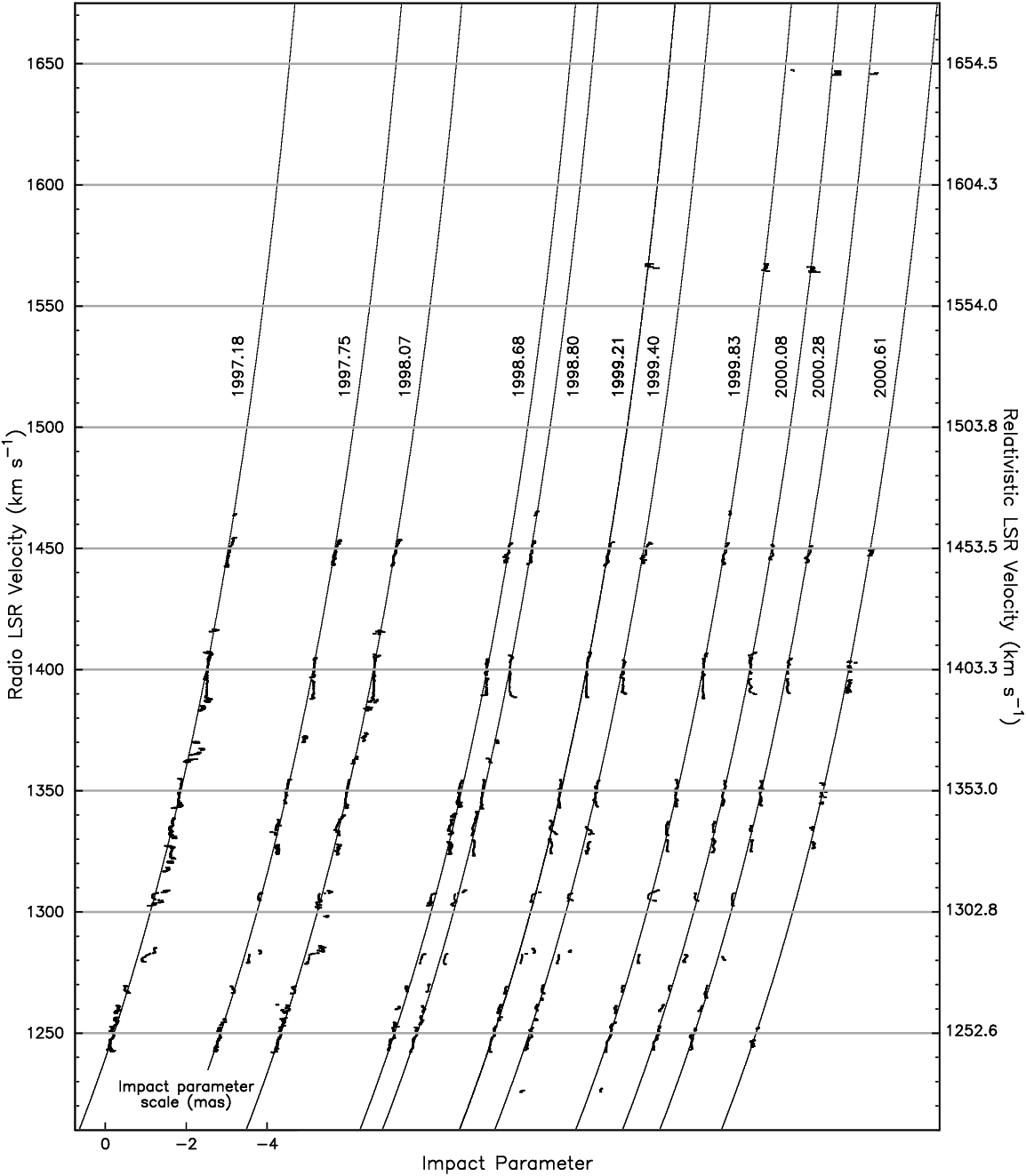}
\caption{Position--velocity traces for high-velocity red-shifted emission, plotted
as in Figure~11. There are fewer traces here because red-shifted emission was only observed
in alternate medium-sensitivity epochs (Table~1).  Keplerian curves (see Figure~10) were 
drawn through the data of each epoch for ease of display.
\label{fig12}}
\end{figure}

\clearpage
                                                                     
\begin{figure}
\includegraphics[scale=0.85]{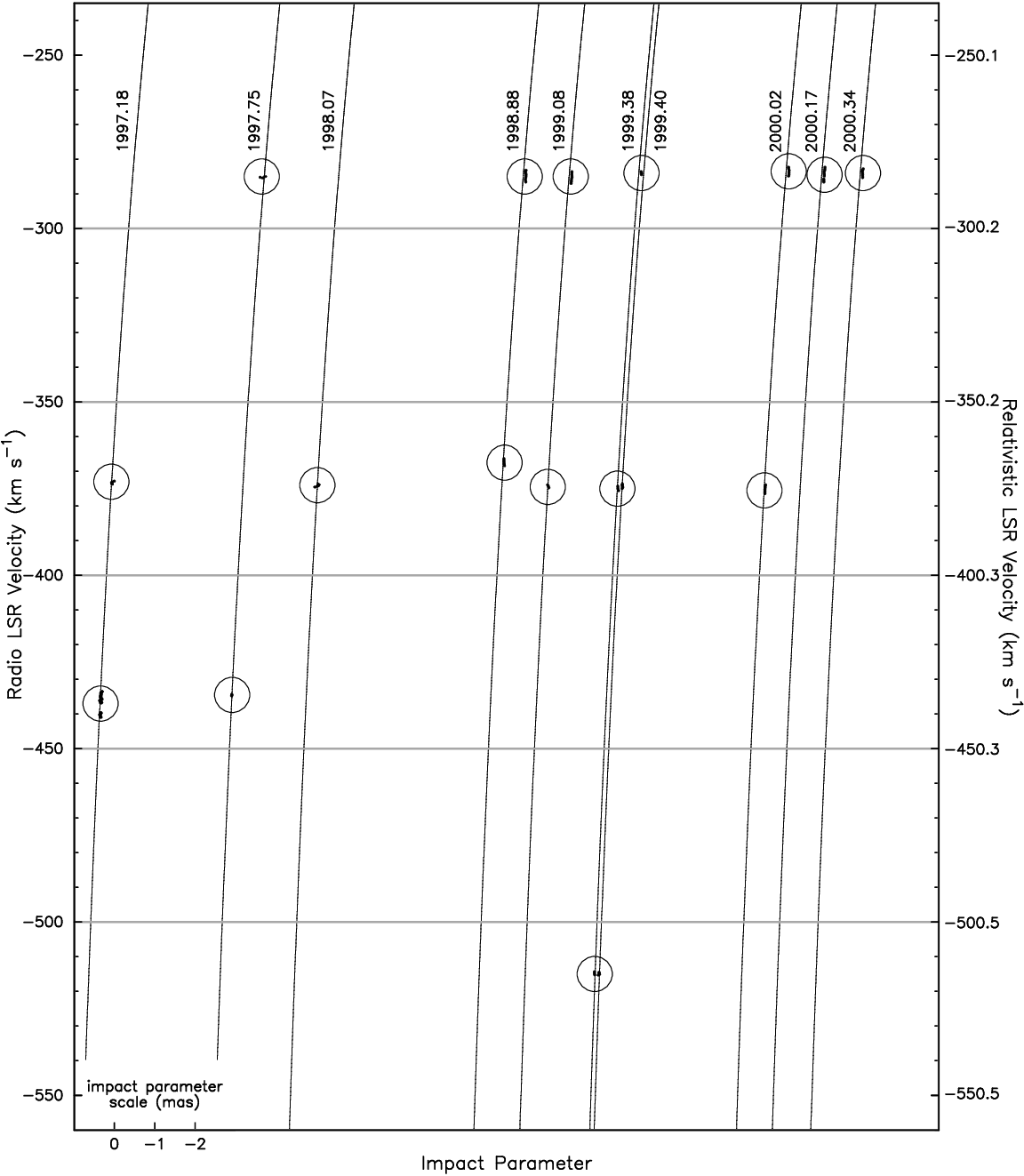}
\caption{Position--velocity traces for high-velocity blue-shifted emission, plotted
as in Figure~11. There are fewer traces here because blue-shifted emission was only observed
in alternate medium-sensitivity epochs (Table~1). Keplerian curves (see Figure~10) were 
drawn through the data of each epoch (circled) for ease of display.
\label{fig13}}
\end{figure}

\clearpage
                                                                     
\begin{figure}
\includegraphics[scale=1.0]{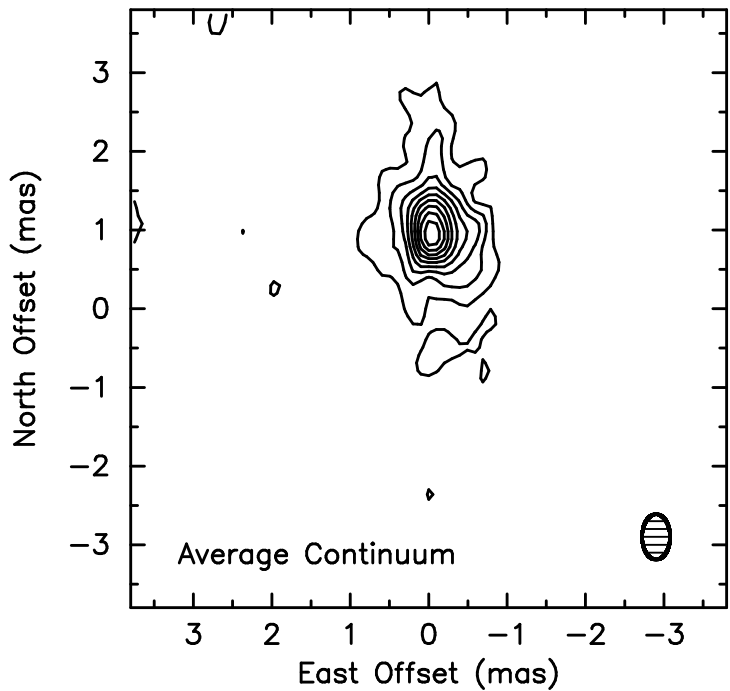}
\caption{Map of time-averaged continuum emission obtained from a uniformly 
weighted average of images from fifteen epochs.  The rms noise level is 0.05 mJy.  
Contours are in $2.5\sigma$ steps, starting at $2.5\sigma$.  An average restoring 
beam is shown in the lower right.  The origin of coordinates is at the position 
of the maser reference feature.
\label{fig14}}
\end{figure}

\clearpage
                                                                     
\begin{figure}
\includegraphics[scale=0.85]{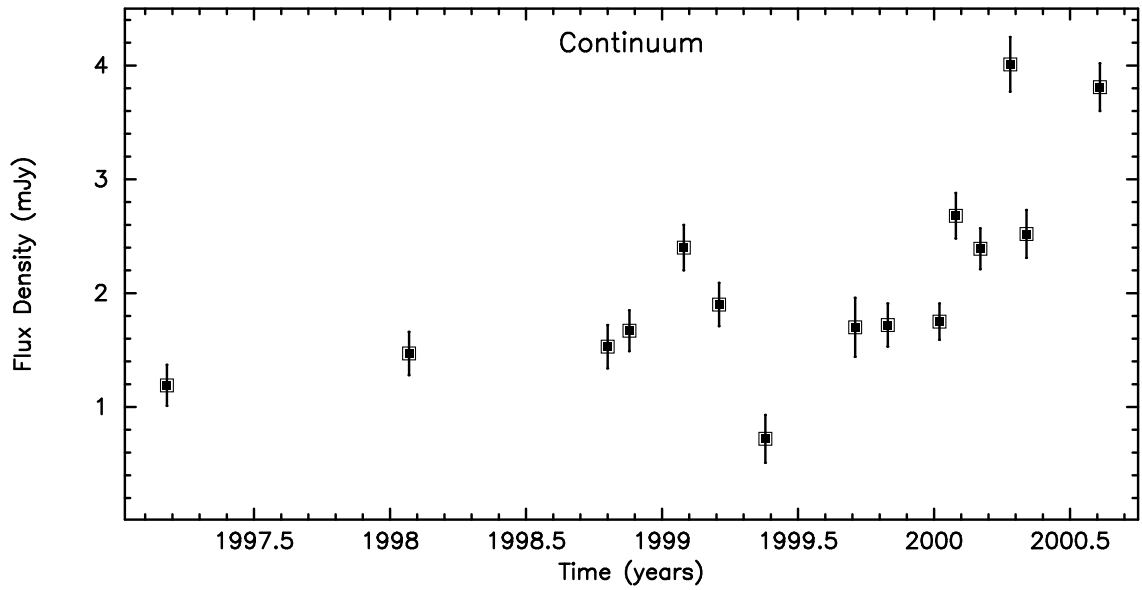}
\caption{Light curve of flux density for epochs with peak continuum emission 
$>3\sigma$ (see Table~6).  Strictly speaking, this is not a total flux density, 
but rather an angle integrated flux density, since we measure it with an 
interferometer that has a limited range of baseline lengths.
\label{fig15}}
\end{figure}

\clearpage
                                                                     
\begin{figure}
\epsscale{1.0}
\plotone{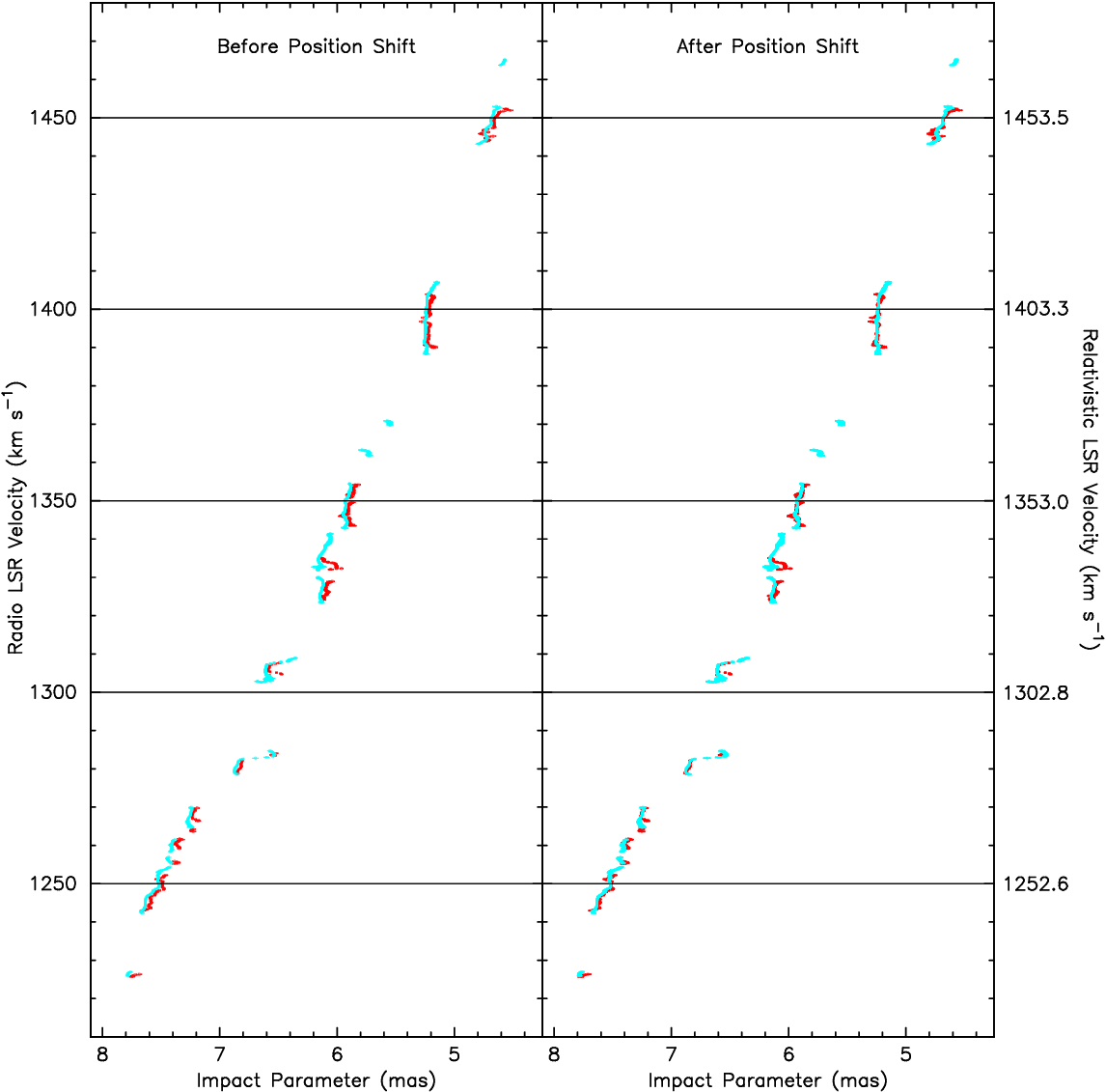}
\caption{Position--velocity traces for red-shifted emission that demonstrate the 
magnitude of the postion offset affecting some epochs and the effectiveness of correction 
($\S3.2$).  Both panels show the weighted average of red-shifted masers in
medium-sensitivity epochs {\it (blue)} and the trace for one high-sensitivity
observation, BM112H {\it (red)}. {\it (left)} --  No position shift to the BM112H data
is applied.  The masers appear to be offset in position from the weighted average. 
{\it (right)} -- Good agreement after a single position shift (applied to the BM112H
data) that minimizes the offset in a least squares sense. \label{fig16} }
\end{figure}

\clearpage







\end{document}